\documentclass[aps,twocolumn,floats,prd,nofootinbib,superscriptaddress,10pt]{revtex4-1}

\usepackage{graphicx,amsmath,amssymb}
\usepackage{hyperref}
\hypersetup{colorlinks=true, linkcolor=red, filecolor=magenta, urlcolor=blue}
\AtBeginDocument{
    \newwrite\bibnotes
    \def\bibnotesext{Notes.bib}
    \immediate\openout\bibnotes=\jobname\bibnotesext
    \immediate\write\bibnotes{@CONTROL{REVTEX41Control}}
    \immediate\write\bibnotes{@CONTROL{
    apsrev41Control,author="08",editor="1",pages="1",title="0",year="1"}}
     \if@filesw
     \immediate\write\@auxout{\string\citation{apsrev41Control}}
    \fi
}

\begin{document}

\title{{\tt21cmFirstCLASS} I. Cosmological tool for $\Lambda$CDM and beyond}

\author{Jordan Flitter}
\email{E-mail: jordanf@post.bgu.ac.il}
\author{Ely D.\ Kovetz}
\affiliation{Physics Department, Ben-Gurion University of the Negev, Beer-Sheva 84105, Israel}

\begin{abstract}
In this work we present {\tt 21cmFirstCLASS}, a modified version of {\tt 21cmFAST}, the most popular code in the literature for computing the inhomogeneities of the 21-cm signal. Our code uses the public cosmic microwave background (CMB) Boltzmann code {\tt CLASS}, to establish consistent initial conditions at recombination for any set of cosmological parameters and evolves them throughout the dark ages, cosmic dawn, the epoch of heating and reionization. We account for inhomogeneity in the temperature and ionization fields throughout the evolution, crucial for a robust calculation of both the global 21-cm signal and its fluctuations. We demonstrate how future measurements of the CMB and the 21-cm signal can be combined and analyzed with {\tt 21cmFirstCLASS} to obtain constraints on both  cosmological and astrophysical parameters and examine degeneracies between them. As an example application, we show how {\tt 21cmFirstCLASS} can be used to study cosmological models that exhibit non-linearities already at the dark ages, such as scattering dark matter (SDM). For the first time, we present self-consistent calculations of the 21-cm power spectrum in the presence of SDM during the non-linear epoch of cosmic dawn. The code is publicly available at \href{https://github.com/jordanflitter/21cmFirstCLASS}{https://github.com/jordanflitter/21cmFirstCLASS}.

\end{abstract}

\maketitle

\section{Introduction}

After a rapid inflationary epoch ended, our Universe continued to expand and cool down. At some point, nearly 400,000 years after the Big Bang, its temperature was low enough that atoms could first form in a key cosmological moment called \emph{recombination}. Meanwhile, collapsing halos of cold dark matter (CDM) provided the first gravitational seeds for galaxy and star formation. Efficient radiation that was emitted from the first stars and their remnants then ionized the surrounding inter-galactic medium (IGM) during the epoch of \emph{reionization} (EoR). Recently on cosmic timescales, the expansion of the Universe has become dominated by the mysterious force of dark energy. This is a brief description of the concordance cosmological model ($\Lambda$CDM)~\cite{Carroll:2000fy, Peebles:1984ge, Peebles:2002gy, Bull:2015stt}.

Many observables support $\Lambda$CDM as being the correct cosmological model of our Universe. Galaxy surveys of our local Universe at redshift $z\lesssim1$~\cite{Wang:2020dtd, Zhao:2021ahg, BOSS:2016off, BOSS:2016teh, eBOSS:2020yzd, 2dFGRSTeam:2002tzq, Blake:2011wn,Johnson:2014kaa,vanUitert:2017ieu, DES:2022tjd, DES:2022qpf, DES:2021lsy, DES:2021bpo, DES:2021wwk, DES:2020cbm, DES:2020mlx}, and measurements of the Ly$\alpha$ forest ($z\lesssim2.5$)~\cite{Croft:1997jf, McDonald:1999dt, SDSS:2004aee, Viel:2004bf, SDSS:2004kjl, SDSS:2004aee} have found very good agreement between the observed spatial distribution of galaxies and the theoretical predicted distribution. But perhaps it was the cosmic microwave background (CMB)~\cite{Dodelson:2003ft, Lesgourgues:2013qba, Hu:1995fq, Challinor:2009tp, Hu:2008hd, 2009AIPC.1132...86C, Hu:1995em, Hu:1997hv, Kosowsky:1998mb, Cabella:2004mk, Lin:2004xy, Lewis:2006fu}---a form of radiation that has been nearly freely propagating since recombination at $z\sim1100$---that gave $\Lambda$CDM its greatest triumph; measurements of the temperature and polarization anisotropies of the CMB, carried out by the Planck satellite~\cite{Planck:2018nkj} and ground based experiments~\cite{SPT-3G:2022hvq,ACT:2023ipp}, allowed to constrain the six parameters of $\Lambda$CDM to a sub-percent precision level~\cite{Planck:2018vyg}.

Despite its success, the $\Lambda$CDM model does suffer from tensions with observations (see recent reviews in Refs.~\cite{Bullock:2017xww,Perivolaropoulos:2021jda,Kamionkowski:2022pkx,Abdalla:2022yfr}), and more cosmological data is required to resolve them, especially in the large volume between $2.5\lesssim z\lesssim1100$ where our Universe has not been systematically mapped yet. Since according to Big-Bang nucleosynthesis (BBN)~\cite{Cyburt:2015mya, Pitrou:2018cgg} the IGM in our Universe is expected to contain huge amounts of neutral hydrogen before reionization, the 21-cm signal, being sourced by hyperfine energy transitions in hydrogen atoms~\cite{Madau:1996cs, Barkana:2000fd, Loeb:2003ya, Bharadwaj:2004it, Furlanetto:2006jb, Pritchard:2011xb, Bera:2022vhw, Shaw:2022fre}, has become an important target for cosmologists.

Nowadays there are ongoing efforts to detect the 21-cm signal by many different collaborations. Some of them focus on detecting the global signal, that is the sky-averaged signal. These include the Experiment to Detect the Global reionization Signature (EDGES)~\cite{Monsalve:2019baw},  Shaped Antenna measurement of the background RAdio Spectrum (SARAS)~\cite{2021arXiv210401756N}, Large-Aperture Experiment to Detect the Dark Ages (LEDA)~\cite{2018MNRAS.478.4193P}, the Radio Experiment for the Analysis of Cosmic Hydrogen (REACH)~\cite{deLeraAcedo:2022kiu} and Probing Radio Intensity at high-Z from Marion (PRIzM)~\cite{2019JAI.....850004P}. In addition, radio interferometer telescopes, such as the Giant Metrewave Radio Telescope (GMRT)~\cite{Pal:2020urw}, the Murchison Widefield Array (MWA)~\cite{Yoshiura:2021yfx}, Low Frequency Array (LOFAR)~\cite{Mertens:2020llj}, the Precision Array for Probing the Epoch of Reionization (PAPER)~\cite{Parsons:2013dwa}, the Hydrogen Epoch of Reionization Array (HERA)~\cite{DeBoer:2016tnn} and the Square Kilometre Array (SKA)~\cite{Braun:2015zta} are devoted to probe the spatial fluctuations in the signal. While most of these experiments are in the stages of noise calibration and have only placed upper bounds on the amplitude of the power spectrum of the signal, the HERA collaboration for example has already extracted a meaningful upper bound on the X-ray luminosity of the first stars~\cite{HERA:2021noe,HERA:2021bsv,HERA:2022wmy} (see also Ref.~\cite{Lazare:2023jkg}).

There are several approaches in the literature for computing the anisotropies in the 21-cm signal. One way is to perform full radiative-transfer hydrodynamic simulations, e.g. {\tt CoDA}~\cite{Ocvirk:2015xzu, Ocvirk:2018pqh,Lewis:2022kwf}, {\tt 21SSD}~\cite{Semelin:2017xgv} and {\tt THESAN}~\cite{Kannan:2021xoz}. Alternatively, post-processing of N-body simulations can be applied with ray-tracing algorithms such as {\tt C$\,^2$-Ray}~\cite{Mellema:2005ht} or {\tt CRASH}~\cite{Maselli:2003ij}. While these simulations improved our understanding of the physics of the EoR and helped to refine reionization models, they are computationally expensive and cannot be used for parameter inference. Faster approximated schemes that solve the one dimensional radiative transfer equation can be found in the codes of {\tt BEARS}~\cite{Thomas:2008uq}, {\tt GRIZZLY}~\cite{Ghara:2017vby} and {\tt BEoRN}~\cite{Schaeffer:2023rsy}. There are also approximated purely analytic prescriptions in the literature, e.g.~\cite{Schneider:2020xmf}. In {\tt Zeus21}~\cite{Munoz:2023kkg} the 21-cm power spectrum at $z\gtrsim10$ can be evaluated in seconds, thanks to an approximated exponential fit that relates the linear matter density fluctuations to the non-linear fluctuations of the star formation rate density (SFRD). Finally, semi-numerical codes that implement the excursion-set formalism~\cite{Furlanetto:2004ha} are widely used in the literature, from~\cite{Reis:2021nqf} to {\tt SimFast21}~\cite{2010ascl.soft10025S} to the ever-popular {\tt 21cmFAST}~\cite{Mesinger:2010ne, Munoz:2021psm}.

In this paper we introduce our code for calculating the 21-cm anisotropies. We call it {\tt 21cmFirstCLASS}. It is essentially the merger of the two well-known codes---{\tt 21cmFAST}\footnote{\href{https://github.com/21cmfast/21cmFAST}{github.com/21cmfast/21cmFAST} (we currently use v.\ 3.3.1)} and the linear Boltzmann solver {\tt CLASS}\footnote{\href{https://github.com/lesgourg/class_public}{github.com/lesgourg/class\_public}}~\cite{Blas:2011rf}. 
In this version, the Lyman-Werner (LW) feedback~\cite{Haiman:1996rc, Bromm:2003vv, Fialkov:2012su} as well as the relative velocity between baryons and CDM ($V_\mathrm{cb}$)~\cite{Tseliakhovich:2010bj, Tseliakhovich:2010yw, Fialkov:2011iw, Ali-Haimoud:2013hpa} are taken into account in each cell, while pop-II and pop-III stars are separated into atomic and molecular cooling galaxies, respectively. In addition, the code contains our past modifications to {\tt 21cmFAST} to incorporate the Ly$\alpha$ heating mechanism~\cite{Venumadhav:2018uwn, Reis:2021nqf,  Sarkar:2022dvl}, as well as the ability to consider fuzzy dark matter (FDM) with an arbitrary mass and fraction~\cite{Flitter:2022pzf}.

There are three main advantages to {\tt 21cmFirstCLASS}: (1) It generates consistent initial conditions (via {\tt CLASS}) and thereby allows one to study degeneracies between cosmological parameters and astrophysical parameters. (2) It allows a combined analysis of CMB and 21-cm anisotropies, which improves constraining power and allows for degeneracy breaking. (3) Unlike the standard {\tt 21cmFAST} code which is designed to begin the simulation at $z=35$ with a homogeneous temperature field, the user can control the initial redshift of {\tt 21cmFirstCLASS}, and even set it to recombination. As a consequence, the fields in the box are evolved non-uniformly from an early redshift, naturally leading to the correct state of the box at $z=35$. This is particularly important for beyond $\Lambda$CDM models which exhibit non-linear fluctuations early on, e.g.\ in scenarios with primordial magnetic fields~\cite{Cruz:2023rmo}. 

To demonstrate the last point, we consider as an example an exotic dark matter model which we refer to as scattering dark matter (SDM). In this model, some part of the dark matter is composed of particles which are able to interact non-gravitationally with ordinary matter and scatter off of it elastically~\cite{Dvorkin:2013cea, Munoz:2015bca, Boddy:2018wzy, Fialkov:2018xre, Xu:2018efh, Barkana:2018lgd, Short:2022bmm, He:2023dbn, Driskell:2022pax, McDermott:2010pa, Kovetz:2018zan, Munoz:2018pzp, Berlin:2018sjs, Barkana:2018qrx, Slatyer:2018aqg, Liu:2018uzy, Munoz:2018jwq, Ali-Haimoud:2015pwa, Gluscevic:2017ywp, Boddy:2018kfv, Maamari:2020aqz, Nguyen:2021cnb, Buen-Abad:2021mvc, Rogers:2021byl, Liu:2019knx, Barkana:2022hko}. In that context, this work resembles the work of Ref.~\cite{Driskell:2022pax}, but there are a few important differences. First, Ref.~\cite{Driskell:2022pax} used {\tt ARES}~\cite{Mirocha:2014faa, 2017MNRAS.464.1365M} in their astrophysical calculations. This code assumes a simpler astrophysical model than {\tt 21cmFAST}; it does not account for halo mass dependence in the calculation of the star formation efficiency and it lacks treatment for star suppression feedbacks in molecular cooling galaxies. Moreover, it computes global astrophysical quantities (e.g.\ emissivity) from global cosmological quantities (e.g.\ halo mass function) and therefore does not take into account important non-linear fluctuations at low redshifts. And secondly, in Ref.~\cite{Driskell:2022pax}, the astrophysical parameters were fixed in the analysis and it focused on the global signal. We on the other hand vary both cosmological and astrophysical parameters and derive forecasts with respect to the 21-cm power spectrum while simulating HERA's noise. We demonstrate that HERA in its design sensitivity is expected to easily probe SDM with cross-sections smaller by  an order of magnitude than e.g.\ forecasted constraints for CMB-S4~\cite{Boddy:2018wzy}.

This is not the first work to consider the 21-cm power spectrum in the presence of SDM. However, it is the first work that computes \emph{consistently} the 21-cm power spectrum in the presence of SDM during the non-linear cosmic dawn epoch. For example, Refs.~\cite{Fialkov:2018xre, Munoz:2018jwq, Barkana:2022hko} have estimated the 21-cm power spectrum by considering only the initial Maxwellian fluctuations in the relative velocity between the SDM and the baryons. In {\tt 21cmFirstCLASS}, non-linear fluctuations in the density and the SFRD fields are automatically captured. In follow-up work~\cite{FlitterSDMForecasts}, we will use {\tt 21cmFirstCLASS} to extend the work of Ref.~\cite{Short:2022bmm}, which focused on the linear dark ages epoch, to make detailed forecasts for constraining SDM at cosmic dawn.

While working on this project, inspired by the work of Ref.~\cite{Munoz:2023kkg} (that introduced the {\tt Zeus21} code), we have also studied in detail the impact of early linear fluctuations on the late non-linear 21-cm power spectrum at low redshifts. The results of that analysis can be found in a companion paper~\cite{Flitter:2023rzv} (hereafter referred to as Paper II).

The remaining parts of this paper are organized as follows. In Sec.~\ref{sec: 21cm theory} we briefly outline the physics of the 21-cm signal. In section~\ref{sec: Initial conditions} we describe the initial conditions used in our code and in Sec.~\ref{sec: Comparison between 21cmFirstCLASS and 21cmFAST} we compare the output of {\tt 21cmFirstCLASS} with {\tt 21cmFAST}. In Sec.~\ref{sec: Combining 21cm and CMB data} we demonstrate how 21-cm and CMB data can be readily combined using {\tt 21cmFirstCLASS} to relax degeneracies between cosmological parameters. We then move on to discuss the SDM physics and its implementation in {\tt 21cmFirstCLASS} in Sec.~\ref{sec: Scattering dark matter}. At the end of that section, the results of the SDM evolution and its impact on the 21-cm power spectrum are presented, as well as forecasts for its detectability by HERA. We provides our conclusions in Sec.~\ref{sec: Conclusions}.

Throughout this paper, we adopt the best-fit values for the cosmological parameters from Planck 2018 ~\cite{Planck:2018vyg} (without BAO), namely we assume a Hubble constant $h=0.6736$, a primordial curvature amplitude $A_s=2.1\times10^{-9}$ with a spectral index $n_s=0.9649$, and total matter and baryons density parameters $\Omega_m=0.3153$, $\Omega_b=0.0493$. For the CMB calculations we also assume an optical depth to reionization $\tau_\mathrm{re}=0.0544$ and a single species of massive neutrinos with mass $m_\nu=0.06\,\mathrm{eV}$. For the fiducial values of the astrophysical parameters in {\tt 21cmFAST} and {\tt 21cmFirstCLASS}, we adopt the EOS2021 values listed in Table 1 of Ref.~\cite{Munoz:2021psm}. All of our formulae are expressed in the CGS unit system. To reduce clutter, we often do not explicitly write the independent arguments of the physical quantities (e.g.\ redshift, wavenumber, etc.) and they should be inferred from the context.

\section{21cm theory}\label{sec: 21cm theory}
The observed physical quantity of the 21-cm signal is known as the brightness temperature, which reflects the excess or deficit of CMB photons at a given frequency (or redshift),
\begin{equation}\label{eq: 1}
T_{21}=\frac{T_s-T_\gamma}{1+z}\left(1-\mathrm{e}^{-\tau_{21}}\right),
\end{equation}
where $T_\gamma\propto\left(1+z\right)$ is the redshift-dependent CMB temperature, $T_s$ is the spin temperature, and $\tau_{21}\ll 1$ is the 21-cm optical depth (see classic reviews of the 21-cm signal in Refs.~\cite{Madau:1996cs, Barkana:2000fd, Bharadwaj:2004it, Furlanetto:2006jb, Pritchard:2011xb}). The spin temperature is a characteristic property of the IGM that measures the relative abundance of hydrogen atoms in the triplet and singlet states, in which the spins of the proton and the electron are aligned and anti-aligned, respectively. As the Universe evolves, various processes excite the hydrogen gas and compete between themselves on setting the value of the spin temperature. In thermal equilibrium the spin temperature reads 
\begin{equation}\label{eq: 2}
T_s^{-1}=\frac{x_\mathrm{CMB} T_\gamma^{-1}+x_\mathrm{coll} T_k^{-1}+\tilde x_\alpha T_\alpha^{-1}}{x_\mathrm{CMB}+x_\mathrm{coll}+\tilde x_\alpha}.
\end{equation}
Here, $T_k$ is the IGM gas kinetic temperature, $T_\alpha\approx T_k$ is the color temperature of Ly$\alpha$ photons, and $x_\mathrm{CMB}=\left(1-\mathrm{e}^{-\tau_{21}}\right)/\tau_{21}\sim1$, $x_\mathrm{coll}$ and $\tilde x_\alpha$ are the CMB~\cite{Venumadhav:2018uwn}, collisional~\cite{Furlanetto:2006jb}, and Ly$\alpha$~\cite{Hirata:2005mz} couplings, respectively.

\begin{figure}[b!]
\includegraphics[width=0.92\columnwidth]{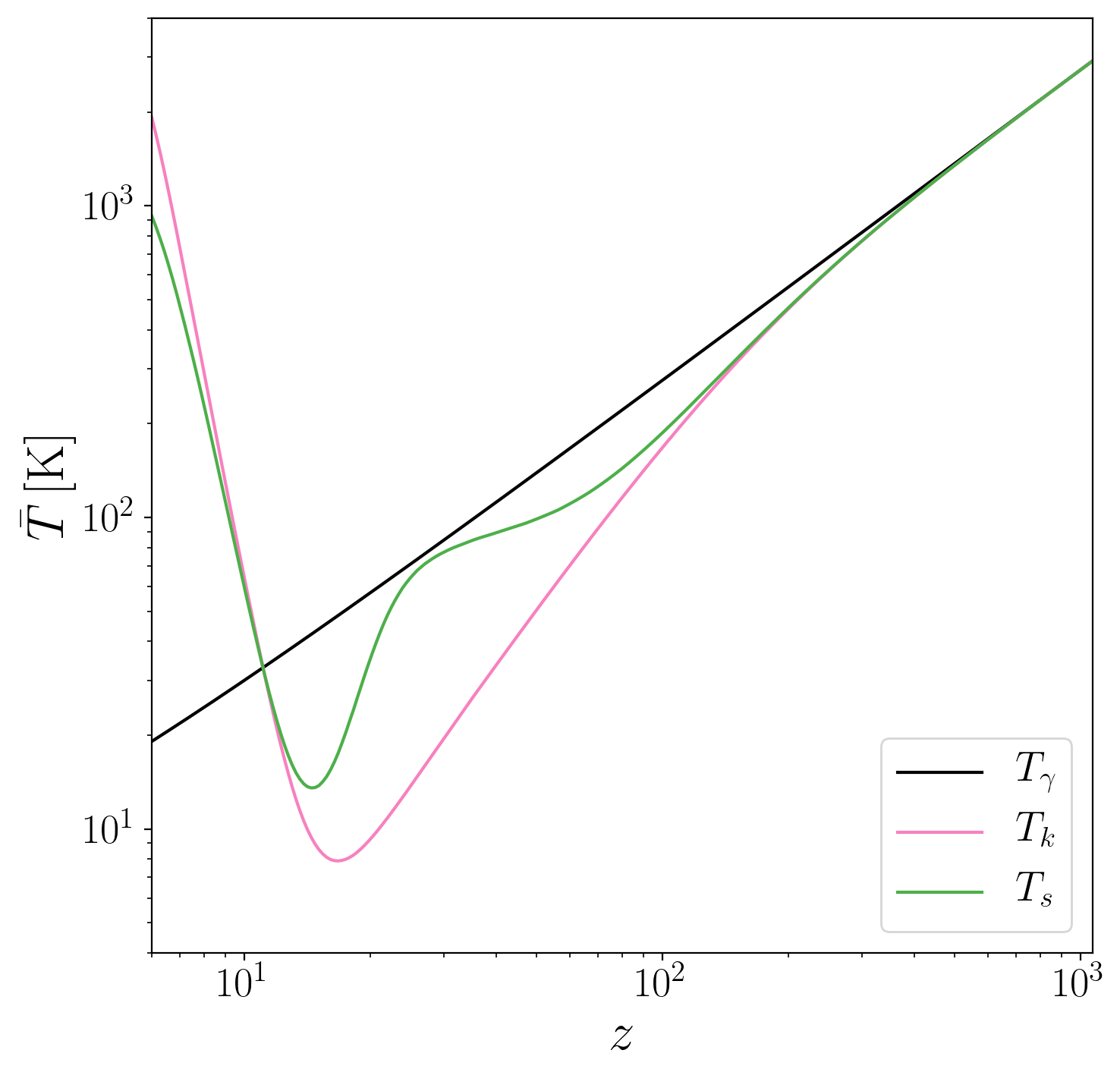}
\caption{The evolution of the globally averaged CMB temperature, gas kinetic temperature and the spin temperature. This figure was made with {\tt 21cmFirstCLASS}.}
\label{Fig: figure_1}
\end{figure}

\begin{figure*}
\includegraphics[width=\textwidth]{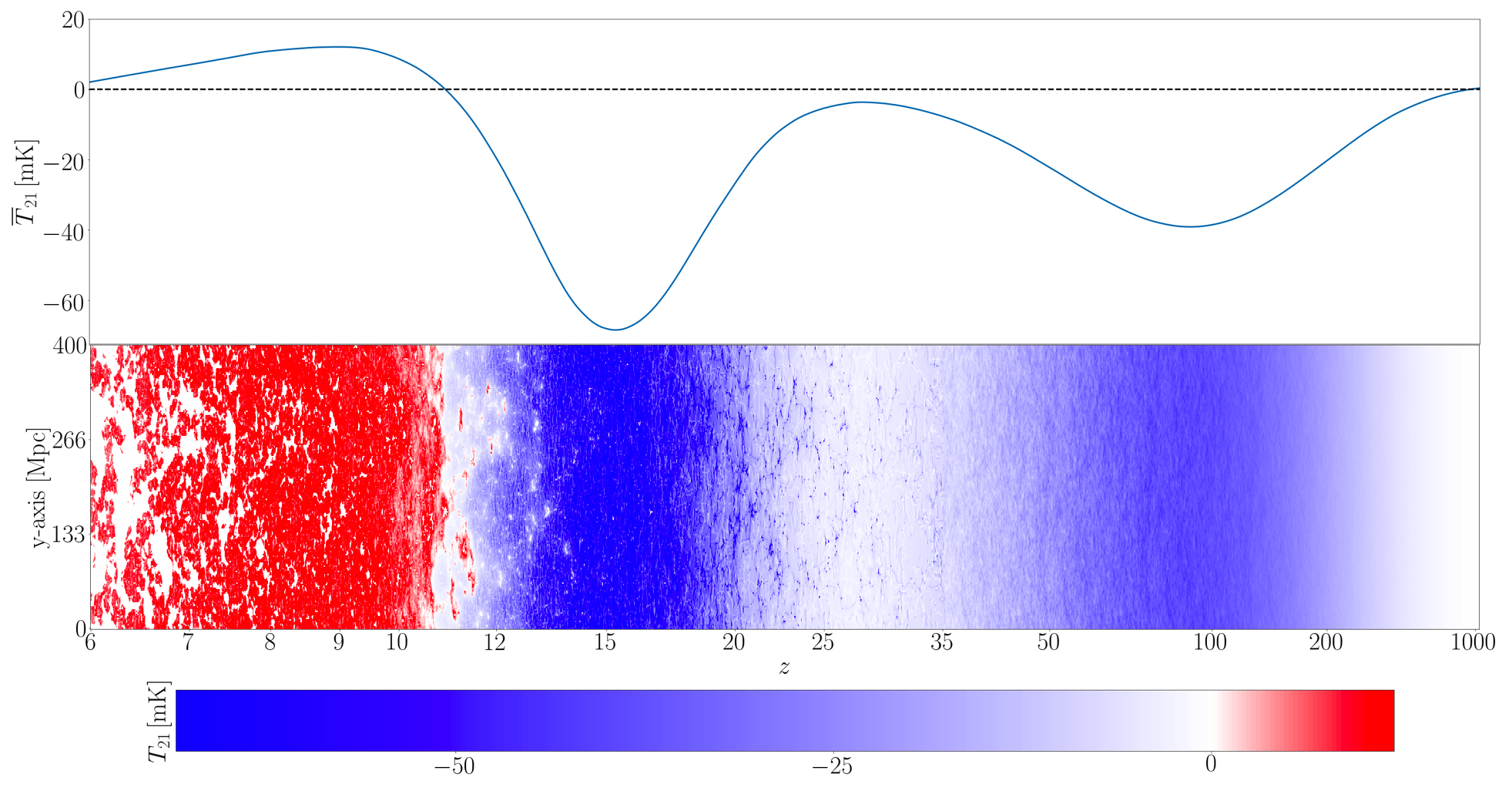}
\caption{\emph{Top panel}: global brightness temperature as a function of redshift. \emph{Botton panel}: fluctuations in the brightness temperature as a function of redshift. Here we present the facet of the lightcone box that is generated by {\tt 21cmFirstCLASS}. For better visulaization, the box that was used for this simulation was of size $400\,\mathrm{Mpc}$ and contained $200^3$ cells. 
The fluctuations pattern seen here is derived from an approximated scale-independent growth, whereas in principle scale-dependent growth should be considered. This assumption will be relaxed in the next version of {\tt 21cmFirstCLASS}. See more details on that point in Paper II.}
\label{Fig: figure_2}
\end{figure*}

As we demonstrate in Fig.~\ref{Fig: figure_1}, the globally averaged value of the spin temperature changes with time. Not long after recombination, at $z\sim1000$, $T_s\approx T_k\approx T_\gamma$. As the Universe expands,  the gas adiabatically cools  and its temperature departs from the CMB temperature, and so the spin temperature settles on an intermediate value, which is determined by the ratio of $x_\mathrm{coll}$ and $x_\mathrm{CMB}$. Since $x_\mathrm{coll}$ is inversely proportional to the volume of the Universe, $x_\mathrm{coll}>1$ at $z\gtrsim100$, and $T_s$ approaches $T_k$. Afterwards, the Universe becomes large enough so that collisional excitations are no longer efficient, $x_\mathrm{coll}<1$, and $T_s$ is driven back towards $T_\gamma$. As can be seen in Fig.~\ref{Fig: figure_2}, the departure of $T_s$ from $T_\gamma$ at $25\lesssim z\lesssim700$ results in the first absorption feature in the globally averaged brightness temperature. This cosmological epoch is known as the \emph{dark ages}. It should be stressed that during this epoch stars have not been formed yet, and therefore the signal is completely determined from cosmology. Hence, within the standard model paradigm, the 21-cm dark ages signal is considered to be well understood theoretically, although it has yet to be measured at that epoch.

The theoretical uncertainty in the 21-cm signal begins after $z\sim25$, once the first stars have been formed. Ly$\alpha$ radiation emitted from the first stars strongly couples the spin temperature back to $T_k$ via the Wouthuysen-Field (WF) effect~\cite{1952AJ.....57R..31W,1958PIRE...46..240F}, and a second absorption feature in the 21-cm signal is expected to be found, although its exact shape and location depend on the assumed astrophysical model and are thus highly uncertain. This epoch is known as  \emph{cosmic dawn}. Depending on the astrophysical parameters, X-rays emitted from stars may heat the surrounding IGM (taking the spin temperature with it) above the CMB temperature, which would lead to an emission feature in the signal. Eventually, stellar radiation reionizes the gas in the IGM and bubbles of ionized hydrogen begin to emerge. After the $\emph{reionization}$ epoch is over, $\tau_{21}\to0$ and the 21-cm signal vanishes.

There are three important quantities which govern the brightness temperature during the cosmic dawn and afterwards. These are the Ly$\alpha$ flux $J_\alpha$ (since $\tilde x_\alpha\propto J_\alpha$), the gas kinetic temperature $T_k$, and the ionization fraction $x_e\equiv n_e/\left(n_\mathrm{H}+n_\mathrm{He}\right)$, where $n_e$, $n_\mathrm{H}$ and $n_\mathrm{He}$, are the free-electron, hydrogen-nuclei and helium-nuclei number-densities, respectively. We will not focus on prescriptions for evaluating $J_\alpha$ in this paper and instead refer the reader to Refs.~\cite{Mesinger:2010ne, Mirocha:2014faa, Hirata:2005mz, Chen:2003gc} for more details. The evolution of $T_k$ is determined from
\begin{equation}\label{eq: 3}
\frac{dT_k}{dz}=\frac{dt}{dz}\left[-2HT_k+\Gamma_C\left(T_\gamma-T_k\right)+\left.\frac{dT_k}{dt}\right|_\mathrm{ext}\right],
\end{equation}
where $dt/dz=-\left[H\left(1+z\right)\right]^{-1}$, $H$ is the Hubble parameter, and $\Gamma_C$ is the Compton heating rate,
\begin{equation}\label{eq: 4}
\Gamma_C\equiv\frac{8\pi^2\sigma_T\left(k_BT_\gamma\right)^4}{45\hbar^3c^4m_e}\frac{x_e}{1+x_e}.
\end{equation}
Here, $c$ is the speed of light, $\hbar$ is the (reduced) Planck constant, $k_B$ is Boltzmann's constant, $m_e$ is the electron mass, and $\sigma_T$ is Thomson cross-section. The term
$dT_k/dt|_\mathrm{ext}$ that appears in Eq.~\eqref{eq: 3} represents the "external" heating rates,
\begin{equation}\label{eq: 5}
\left.\frac{dT_k}{dt}\right|_\mathrm{ext}=\epsilon_\mathrm{ext}+\frac{2}{3}\frac{T_k}{1+\delta_b}\frac{d\delta_b}{dt}-\frac{T_k}{1+x_e}\frac{dx_e}{dt},
\end{equation}
where $\epsilon_\mathrm{ext}$ denotes the heating rates that come from external sources, mainly X-ray heating (but Ly$\alpha$ as well as CMB heating rates~\cite{Reis:2021nqf, Venumadhav:2018uwn, Sarkar:2022dvl} can be included), and $\delta_b\equiv\delta\rho_b/\bar\rho_b$ is the contrast in the baryon-density fluctuations. The reason for  why we classify the last two terms in Eq.~\eqref{eq: 5} as ``external" heating rates, even though they are sourced by the adiabatic cooling of the IGM, will become clear in Appendices~\ref{sec: Compton tight coupling approximation} and \ref{sec: Dark matter tight coupling approximation} where we derive the tight coupling approximations.

From Eqs.~\eqref{eq: 3}-\eqref{eq: 5} it can be seen that the evolution of $T_k$ depends on $x_e$, especially at early times when the Compton heating rate dominates. The exact detailed evolution of $x_e$ at early times on the other hand is quite intricate as it requires tracking the recombination states of both hydrogen and helium, while taking into account excitations to high-order energy levels. In the seminal work of Refs.~\cite{Ali-Haimoud:2010hou, Lee:2020obi}, these effects have been shown to have a sub-percent impact on the evolution of $x_e$, making them crucial for analyzing the CMB anisotropies at the precision level of the Planck satellite data. A  state-of-the-art recombination code that implements these effects and is publicly available is {\tt HyRec}\footnote{\href{https://github.com/nanoomlee/HYREC-2}{github.com/nanoomlee/HYREC-2}}, which we have incorporated in our {\tt 21cmFirstCLASS} code.

Yet, in order to derive the evolution of temperature and ionization fluctuations within a few percent error, we show in Paper II that it is sufficient to consider Peebles' effective three-level atom model~\cite{1968ApJ...153....1P}, in which the evolution of $x_e$ reads
\begin{equation}\label{eq: 6}
\frac{dx_e}{dz}=\frac{dt}{dz}\left[\left.\frac{dx_e}{dt}\right|_\mathrm{reio}+\mathcal C\left(\beta_\mathrm{ion}\left(1-x_e\right)-\alpha_\mathrm{rec}n_\mathrm{H}x_e^2\right)\right],
\end{equation}
where $\alpha_\mathrm{rec}$ is the recombination rate (in units of $\mathrm{cm^3/sec}$), $\beta_\mathrm{ion}$ is the early photoionization rate, and $\mathcal C$ is the Peebles coefficient (see Appendix A in Paper II for more details on the Peebles coefficient). The term $dx_e/dt|_\mathrm{reio}$ denotes the reionization rate at late times. At early times (long before reionization started), the recombination rate and photoionization rates were in equilibrium, implying that
\begin{equation}\label{eq: 7}
\beta_\mathrm{ion}=\alpha_\mathrm{rec}\left(\frac{m_ek_BT_\gamma}{2\pi\hbar^2}\right)^{3/2}\mathrm{e}^{-\epsilon_0/\left(k_BT_\gamma\right)},
\end{equation}
where $\epsilon_0=13.6\,\mathrm{eV}$ is the ionization energy of the hydrogen atom from its ground state.

Since the standard {\tt 21cmFAST} code begins its calculations at $z=35$, the $\beta_\mathrm{ion}$ term is completely negligible and was omitted. The recombination rate is the case-A recombination rate which accounts for recombination to the ground state~\cite{Abel:1996kh}. In addition, the factor $\mathcal C$ is not interpreted as the Peebles coefficient but rather as the clumping factor $\langle x_e^2\rangle/\langle x_e\rangle^2$~\cite{Mao:2019vob}, which the code sets as a  constant with a value of $2$ to account for unresolved sub-grid fluctuations. This serves as an excellent approximation to the evolution of $x_e$ at late times. We have confirmed with {\tt 21cmFirstCLASS} that using {\tt HyRec} at all redshifts almost replicates the same $x_e$ evolution at low redshifts\footnote{The incorporation of {\tt HyRec} was done only at the code section of {\tt 21cmFAST} that evolves $x_e$, while leaving the reionization code unchanged. Therefore, the reionization history remains almost the same in {\tt 21cmFAST} and {\tt 21cmFirstCLASS}---see more in Sec.~\ref{sec: Comparison between 21cmFirstCLASS and 21cmFAST}.}, while not introducing errors in the 21-cm power spectrum that are larger than HERA's sensitivity (see for example Fig.~\ref{Fig: figure_3} and further discussion in Sec.~\ref{sec: Comparison between 21cmFirstCLASS and 21cmFAST}).

\section{Initial conditions}\label{sec: Initial conditions}

Our code, {\tt 21cmFirstCLASS}, is composed of two main codes: (1) {\tt CLASS}, which generates the consistent initial conditions at recombination, and (2) a modification of  {\tt 21cmFAST}, which uses the  initial conditions from {\tt CLASS} to generate the initial box and then evolves this box until the 21-cm signal vanishes. In the remaining parts of this paper, we use a box of comoving size $256\,\mathrm{Mpc}$ and a resolution\footnote{To be perfectly clear, by 128 we refer to the parameter {\tt HII\_\,DIM}, and not the parameter {\tt DIM}, which is three times larger.} of $128^3$ cells, and initialize the evolution at recombination. We have confirmed that increasing these specifications does not alter the 21-cm power spectrum beyond HERA's sensitivity.

\subsection{\tt CLASS}

In the standard {\tt 21cmFAST}, the user can vary the cosmological parameters fairly easily from the python wrapper. The varied parameters however only enter in the C-code, while the initial conditions for the simulation remain the same, regardless  the values of the cosmological parameters that were set by the user. This property of {\tt 21cmFAST} makes it inadequate for studying degeneracies between the cosmological parameters and the astrophysical parameters, especially if physics beyond the standard model is considered~(see Sec.~\ref{sec: Scattering dark matter}). In our code, {\tt 21cmFirstCLASS}, the initial conditions for the simulation are completely consistent with the input set of cosmological parameters.

To get the correct initial conditions we use {\tt CLASS}. We allow the user to work with either the primordial curvature amplitude $A_s$ (which is commonly used in the CMB and inflation communities), or with the standard {\tt 21cmFAST} $\sigma_8$ parameter, the matter-density variance, smoothed on a sphere of radius $R_8=8\,h^{-1}\,\mathrm{Mpc}$. Given the current matter-density transfer function $\mathcal T_m\left(k,z=0\right)$, which is one of the outputs of {\tt CLASS}, they are related by 
\begin{equation}\label{eq: 8}
\sigma_8^2=A_s\int_0^\infty\frac{dk}{k}\left(\frac{k}{k_\star}\right)^{n_s-1}W^2\left(kR_8\right)\mathcal T_m^2\left(k,z=0\right),
\end{equation}
where $k_\star=0.05\,\mathrm{Mpc}^{-1}$ is the CMB pivot scale and $W\left(kR_8\right)=3\left(kR_8\right)^{-3}\left[\sin\left(kR_8\right)-kR_8\cos\left(kR_8\right)\right]$ is the Fourier transform of a top-hat filter of radius $R_8$. We note that in $\Lambda$CDM simulations we run {\tt CLASS} with a high $k_\mathrm{max}=4000\,\mathrm{Mpc}^{-1}$, which is necessary to get the correct $\sigma\left(R\right)$ at the relevant scales for {\tt 21cmFAST}.
{\tt CLASS} also computes the background quantities $\bar T_k\left(z\right)$, and $\bar x_e\left(z\right)$, the latter via {\tt HyRec}\footnote{Care has to be taken when converting from {\tt CLASS} conventions for $x_e$, which is $n_e/n_\mathrm{H}$, to {\tt 21cmFAST} conventions, for which $x_e\equiv n_e/\left(n_\mathrm{H}+n_\mathrm{He}\right)$.}. We then define the moment of recombination, and the starting point of our simulation, to be the redshift that solves $x_e\left(z_\mathrm{rec}\right)\equiv0.1$. For the fiducial set of cosmological parameters we use, it is $z_\mathrm{rec}\approx1069$.

In addition, we also evaluate $\mathcal T_{v_\mathrm{cb}}\left(k,z_\mathrm{rec}\right)$, the transfer function of $V_\mathrm{cb}$ during recombination, with
\begin{equation}\label{eq: 9}
\mathcal T_{v_\mathrm{cb}}\left(k,z_\mathrm{rec}\right)=\left|\frac{\theta_c\left(k,z_\mathrm{rec}\right)-\theta_b\left(k,z_\mathrm{rec}\right)}{kc}\right|,
\end{equation}
where $\theta_c$ ($\theta_b$) is the Fourier transform of the divergence of the CDM (baryons) velocity, quantities that are also given by {\tt CLASS}. We construct interpolation tables for $\mathcal T_m\left(k,z=0\right)$, $\mathcal T_{v_{cb}}\left(k,z_\mathrm{rec}\right)$, $\bar T_k\left(z\right)$ and $\bar x_e\left(z\right)$, and they are then used to replace the default tables used by {\tt 21cmFAST}.

Finally\footnote{We also adopt from {\tt CLASS} the helium mass-fraction $Y_\mathrm{He}$ as well as the mean of the relative velocity between CDM and baryons during recombination $\langle V_\mathrm{cb}\left(z_\mathrm{rec}\right)\rangle$. In addition, our code also has the ability to compute the fitting parameters $A_p$, $k_p$ and $\sigma_p$ on the fly (see more details in Eq.~(14) of Ref.~\cite{Munoz:2021psm}), though we find that their effect on the brightness temperature is negligible compared to the effect that $V_{cb}$ has on the minimum halo mass that can still host stars---see Sec.~\ref{sec: Small velocity corrections}.}, we also save {\tt CLASS}'s scale-independent growth factor $D\left(z\right)$ in a new interpolation table that goes into {\tt 21cmFAST}. This quantity is obtained in {\tt CLASS} by solving a second order differential equation. To avoid solving this equation for $D\left(z\right)$, in the standard {\tt 21cmFAST} the ``Dicke growth" factor is used~\cite{Liddle:1995pd, Carroll:1991mt}. This is an analytical fit to the growth factor that works particularly well below $z=35$. However, this fit under-estimates the true growth factor at $z\gtrsim100$, and that can lead to a few percent error in the fluctuations pattern of $T_k$ and $x_e$ at low redshifts. Moreover, these errors can propagate to the global signal at $10\lesssim z\lesssim 20$, when non-linearities become important. To avoid introducing errors in the calculation, without sacrificing runtime or computational cost, we adopt {\tt CLASS}'s growth factor in {\tt 21cmFirstCLASS}. For more details on the scale-independent growth factor and its effect on the 21-cm signal, see Appendix \ref{sec: Scale-independent growth factor}.

\subsection{\tt 21cmFAST}\label{subsec: 21cmFAST}

The initial density and velocity boxes in {\tt 21cmFirstCLASS} are generated in a similar manner as in the standard {\tt 21cmFAST}. Prior to $z=35$, we evolve the matter-density fluctuations linearly\footnote{In this paper we assume $\delta_b\left(z\right)=D\left(z\right)\delta_b\left(z=0\right)$. We comment that although such a scale-independent growth of $\delta_b$ is inadequate at high redshifts, our conclusions in this paper are not affected by this crude assumption, which shall be relaxed in the version of {\tt 21cmFirstCLASS} that will soon be made public. We elaborate much more on this subtlety in Paper II.}, though we have confirmed that evolving the density box non-linearly at high redshifts yields the same 21-cm power spectrum at low redshifts. As for the initial $T_k\left(z_\mathrm{rec}\right)$ and $x_e\left(z_\mathrm{rec}\right)$ boxes, we assume that they are  homogeneous. As we discuss in Paper II, an homogeneous $T_k\left(z_\mathrm{rec}\right)$ box is an excellent assumption, much more than the homogeneous $T_k\left(z=35\right)$ box that is assumed in the standard {\tt 21cmFAST}. For the $x_e$ box, the assumption of homogeneity at $z_\mathrm{rec}\approx1069$ is not justified (though it is still better than assuming homogeneity at $z=35$), but we show in Paper II that the resulting 21-cm power spectrum is not very sensitive to this assumption. Ideally, one could use the $T_k$ and $x_e$ transfer functions from {\tt CLASS} to draw the initial boxes, as was done in Ref.~\cite{Short:2022bmm}. In $\Lambda$CDM, such an approach would remove the necessity of starting the simulation at recombination, since all the fluctuations prior to $z=35$ are linear to a very good approximation. Yet, we stress that this approach is no longer valid in some beyond $\Lambda$CDM cosmologies (like the one we discuss in Sec.~\ref{sec: Scattering dark matter}) where non-linearities have an important role even before $z=35$.

We then solve numerically the differential equation for $T_k$ (Eq.~\eqref{eq: 3}) at each cell, using the Euler method, to promote $T_k$ to the next redshift step, as in the standard {\tt 21cmFAST}. The difference, though, is the step-size. In {\tt 21cmFAST}, a logarithmic redshift sampling is used such that $\left(1+z_{n}\right)/\left(1+z_{n+1}\right)=1.02$, where $z_n$ is the $n$'th redshift sample in the simulation, such that the step-size $\Delta z_n=z_n-z_{n+1}$ is $\sim0.1$ at $z=6$ and $\sim0.6$ at $z=35$. This redshift sampling scheme is insufficient at higher redshifts, and we therefore work with a constant step-size of $\Delta z_n=0.1$ at $35\leq z\leq980$. Above $z=980$ this step-size is also not enough, and we have to switch to $\Delta z_n=0.01$ to simulate the evolution precisely. This fine redshift sampling above $z=980$ comes with a price; although no computationally expensive astrophysical calculations are required, the many redshift samples extend the runtime of the code considerably. Yet, there is a much more clever way to evolve $T_k$ above $z=980$ with  excellent precision, without generating so many redshift samples. Briefly, the idea is to treat the baryons and the CMB as a single fluid when the conditions for the \emph{Compton-TCA} (tight coupling approximation) are satisfied. We provide more details on that method in Appendix~\ref{sec: Compton tight coupling approximation}.

The normal evolution of $x_e$ in {\tt 21cmFirstCLASS} is done with {\tt HyRec}, though our code can be configured to solve instead the Peebles model, Eq.~\eqref{eq: 6}, with the recombination rate $\alpha_\mathrm{rec}=\alpha_B$ of {\tt RECFAST}~\cite{Seager:1999bc, Seager:1999km}, where $\alpha_B$ is the case-B recombination rate (which accounts for recombination only to the first excited state). As in {\tt CLASS}, we use the default ``SWIFT" model of {\tt HyRec} when $T_k/T_\gamma<0.1$, and otherwise we use its ``PEEBLES" model, which is quite similar to  Eq.~\eqref{eq: 6} above. In {\tt CLASS}, however, two quantities are solved with {\tt HyRec}, these are $x_\mathrm{H}$ and $x_\mathrm{He}$. Their relation to $x_e$ is $x_e=x_\mathrm{H}+\left(n_\mathrm{He}/n_\mathrm{H}\right)x_\mathrm{He}$. From this equation the physical meaning of $x_\mathrm{H}$ ($x_\mathrm{He}$) should be clear---it is the contribution of ionized hydrogen (helium) number-density to the total free-electron number-density. In {\tt 21cmFirstCLASS} we assume $x_e\approx x_\mathrm{H}$, which is justified because: (1) helium recombination is over long before hydrogen recombination begins, at $z\sim1500$; (2) the freezout value of $x_\mathrm{He}$ is an order of magnitude smaller than the freezout value of $x_\mathrm{H}$; and (3) the contribution of $x_\mathrm{He}$ to $x_e$ is suppressed by the factor $n_\mathrm{He}/n_\mathrm{H}\approx0.08$. As can be seen in Fig.~\ref{Fig: figure_3}, the assumption $x_e\approx x_\mathrm{H}$ is indeed an excellent approximation.

\section{Comparing {\tt 21\lowercase{cm}F\lowercase{irst}CLASS} with {\tt 21\lowercase{cm}Fast}}\label{sec: Comparison between 21cmFirstCLASS and 21cmFAST}

\begin{figure}
\includegraphics[width=\columnwidth]{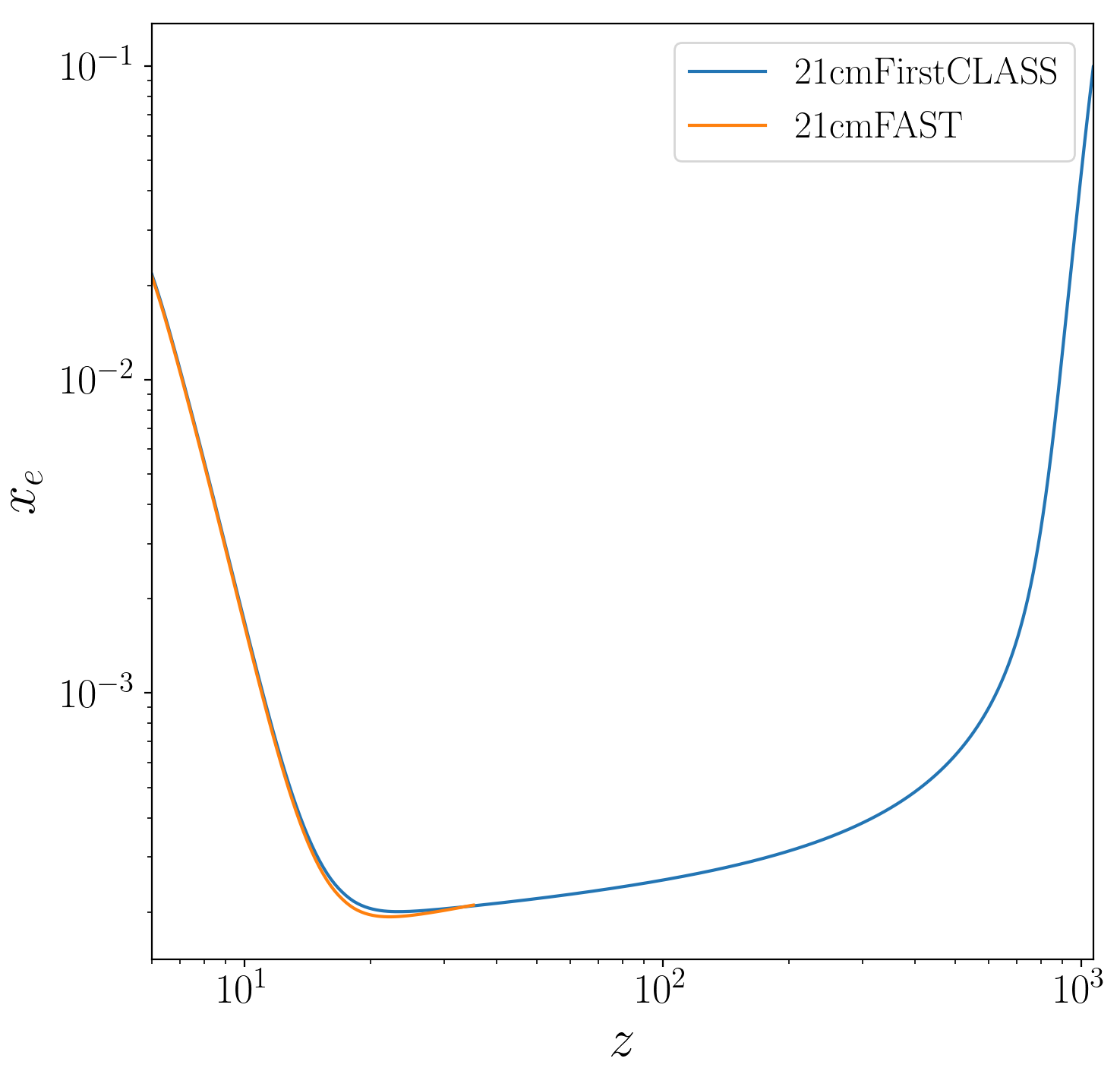}
\caption{Comparison of the global $x_e$ evolution between {\tt 21cmFirstCLASS} and {\tt 21cmFAST} (in both cases we use {\tt CLASS} to obtain the correct initial conditions). In the former, {\tt HyRec} is used all the way from recombination to $z=6$. In the latter, the simulation begins at $z=35$ and $x_e$ is evolved differently (see more details at the end of Sec.~\ref{sec: 21cm theory}). Note the consistency at $z=35$ (though early ionization fluctuations slightly change  the mean of the box in {\tt 21cmFirstCLASS}---see more details in Paper II).}
\label{Fig: figure_3}
\end{figure}

In $\Lambda$CDM, all fluctuations at the relevant scales prior to $z\!=\!35$ can be considered linear to a very good approximation. Consistency therefore implies that {\tt 21cmFirstCLASS} must be able to generate the same initial conditions as in {\tt 21cmFAST}, at $z\!=\!35$. Such a sanity check is demonstrated in Fig.~\ref{Fig: figure_3}, where we present the evolution of $\bar x_e$ in the two codes. At $z\!=\!35$ the two codes agree. Afterwards, the solution of the two codes deviates because of the different evolution, as was outlined at the end of Sec.~\ref{sec: 21cm theory}. This leads to a maximum 5\% difference.

\begin{figure}
\includegraphics[width=\columnwidth]{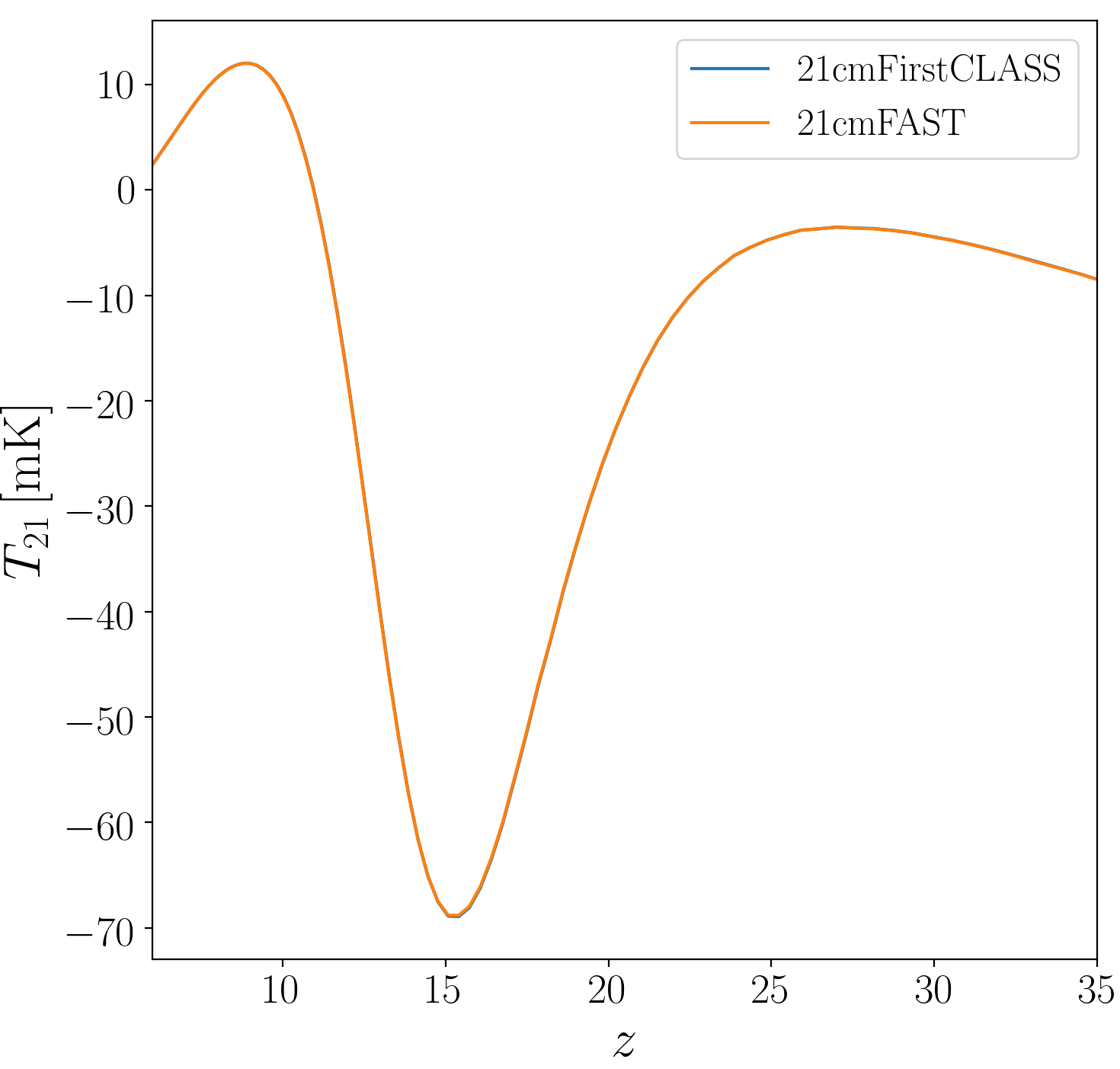}
\caption{Comparison of the global brightness temperature between {\tt 21cmFirstCLASS} and {\tt 21cmFAST}. The curves almost totally overlap.}
\label{Fig: figure_4}
\end{figure}

Yet, this error does not propagate to the observable---the brightness temperature---as can be seen in Fig.~\ref{Fig: figure_4}. This is because $\tau_{21}$ is not proportional to $x_e$ but rather to $x_\mathrm{HI}$, the neutral hydrogen fraction. Before the onset of reionization, we can approximate $x_\mathrm{HI}\approx1-x_e$, and a simple calculation shows that the 5\% difference in $\bar x_e$ translates to merely a 0.001\% error in $\bar x_\mathrm{HI}$.

Even though the first-order statistics of the box, namely its mean, is consistent in both codes, it does not imply that higher-order statistics, like the two-point function, are the same. The Fourier transform of the two-point correlation function is the power spectrum. For the 21-cm signal, it is customary to work with a power spectrum that has  units of $\mathrm{mK^2}$,
\begin{equation}\label{eq: 10}
\Delta_{21}^2\left(k,z\right)=\frac{k^3\bar T_{21}^2\left(z\right)P_{21}\left(k,z\right)}{2\pi^2},
\end{equation}
where $\bar T_{21}$ is the global brightness temperature and $P_{21}\left(k,z\right)$ is the angle-averaged Fourier transform of the two-point function $\langle\delta_{21}\left(\mathbf x,z\right)\delta_{21}\left(\mathbf x',z\right)\rangle$, while $\delta_{21}$ is the local contrast in the brightness temperature, $\delta_{21}\left(\mathbf x,z\right)\equiv T_{21}\left(\mathbf x,z\right)/\bar T_{21}\left(z\right)-1$. We use the {\tt powerbox}\footnote{\href{https://github.com/steven-murray/powerbox}{github.com/steven-murray/powerbox}} package~\cite{2018JOSS....3..850M} to compute $\Delta_{21}^2\left(k,z\right)$ from chunks of the lightcone box of {\tt 21cmFirstCLASS}.

\begin{figure}
\includegraphics[width=0.9\columnwidth]{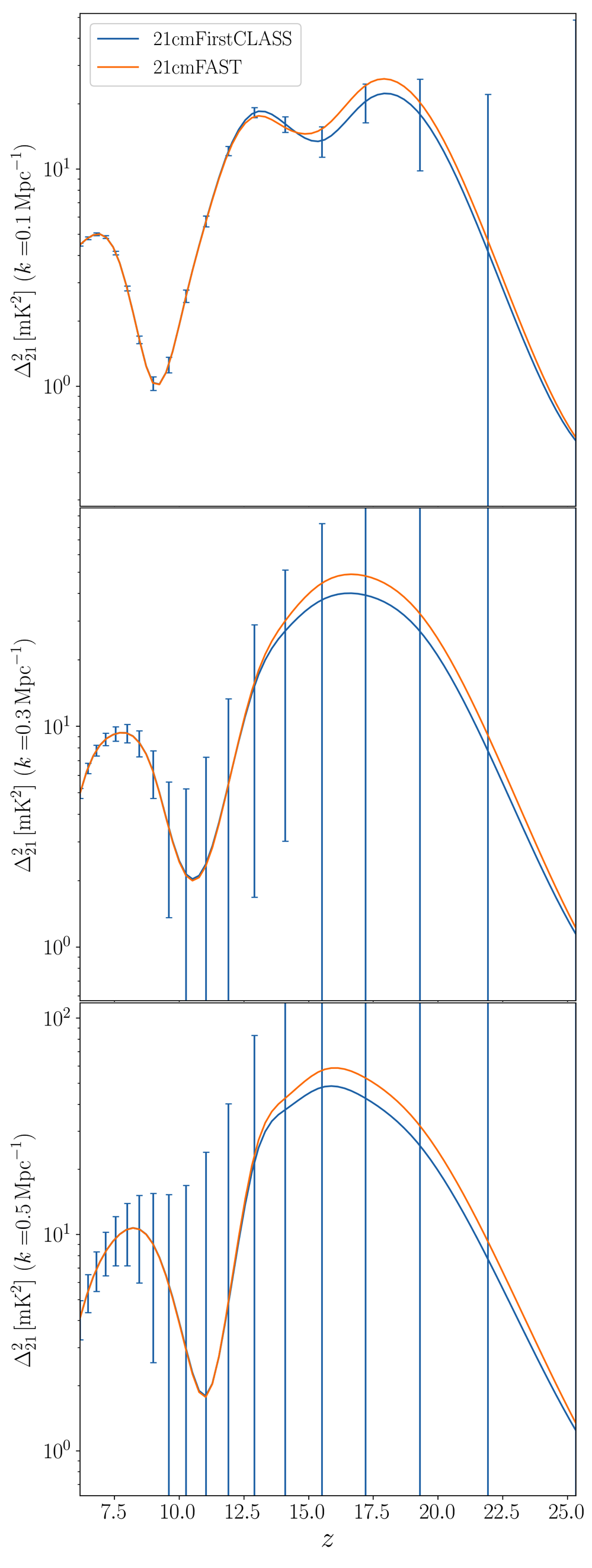}
\caption{Comparison of the 21-cm power spectrum between {\tt 21cmFirstCLASS} and {\tt 21cmFAST}, for three different wavenumbers. The error bars correspond to HERA's noise in its design sensitivity under the assumption of optimistic foregrounds (see Sec.~\ref{sec: Combining 21cm and CMB data}). As we explain in the text, the source for the differences between the curves is early temperature and ionization fluctuations---see more details in Paper II.}
\label{Fig: figure_5}
\end{figure}

In Fig.~\ref{Fig: figure_5}, we compare the 21-cm power spectrum of {\tt 21cmFirstCLASS} and {\tt 21cmFAST}. Unlike the global signal, clear differences can be seen---only because we started the simulation at a different initial time (recombination in {\tt 21cmFirstCLASS} and $z=35$ in {\tt 21cmFAST}). The origin of this effect comes from \emph{early temperature and ionization fluctuations}. The impact of the former kind of fluctuations---temperature fluctuations---on the 21-cm power spectrum was first discussed in Ref.~\cite{Munoz:2023kkg}. In Paper II we extend the discussion on early fluctuations and explore in great detail the contribution of both temperature and ionization fluctuations. Still, Fig.~\ref{Fig: figure_5} suggests that taking into account early temperature and ionization fluctuations results in a maximum distortion of $\sim\!20\%$ to the 21-cm power spectrum at $k=0.3\,\mathrm{Mpc}^{-1}$, $z=17$, which is below HERA's noise level. We note that this is in slight contrast with the conclusions of Ref.~\cite{Munoz:2023kkg}, where larger deviations are claimed. Again, we refer the reader to Paper II for an elaborate discussion on that point.

\section{Combining 21cm and CMB data}\label{sec: Combining 21cm and CMB data}

The anisotropies in the CMB have proven to be an invaluable source for studying different cosmological models. Likewise, the global 21-cm signal and its inhomogeneities are expected to contain rich information that can be employed in cosmological studies of $\Lambda$CDM and beyond. Because {\tt 21cmFirstCLASS} already calculates the CMB anisotropies via {\tt CLASS}, it is only natural to include them as part of our analysis. These two observables are uncorrelated and thus can be used to break degeneracies in the other observable. We will demonstrate this point below while working with the Fisher formalism.

\subsection{Fisher Formalism}\label{sec: Fisher formalism}

\begin{figure*}
\includegraphics[width=0.7\textwidth]{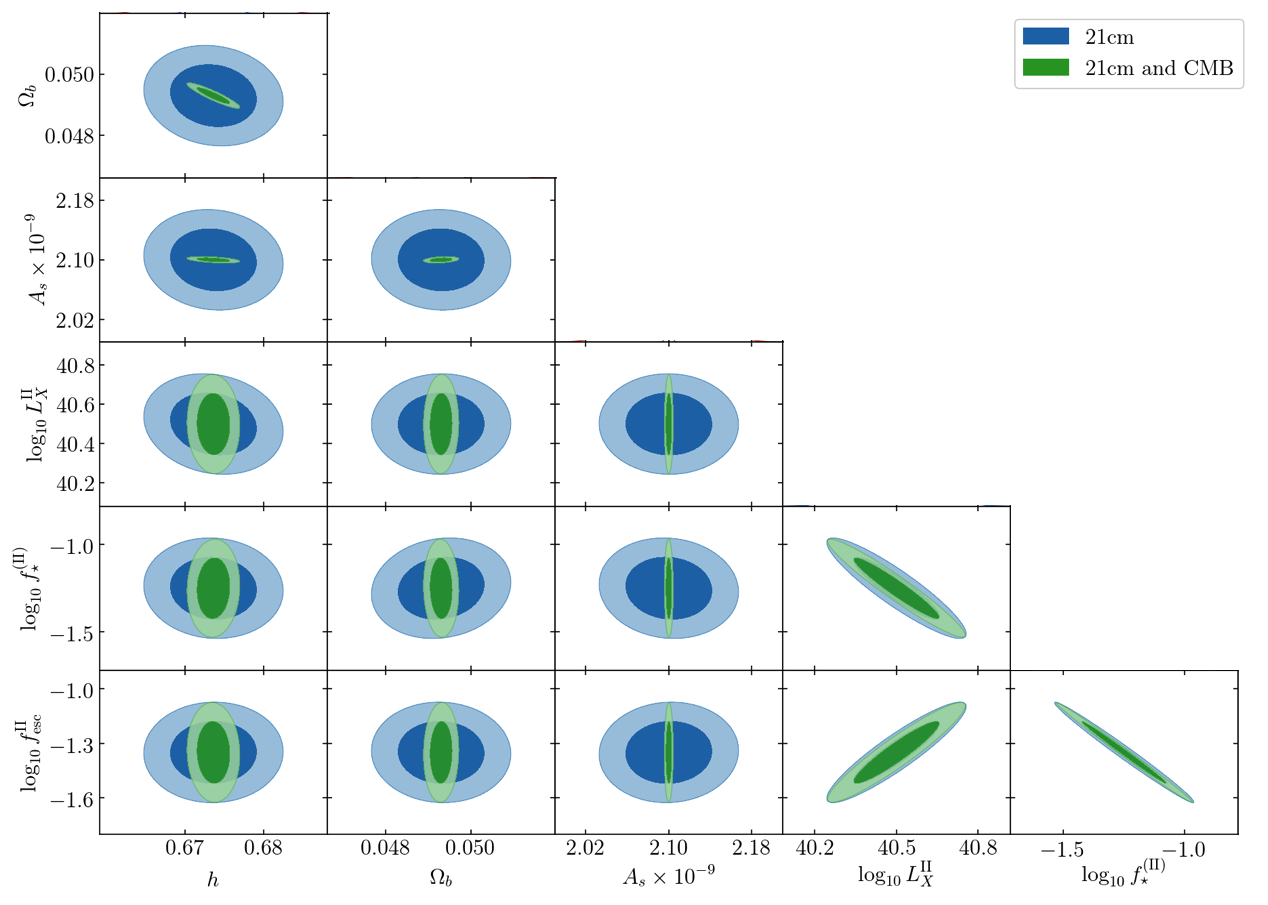}
\caption{Forecasts of 1-$\sigma$ and 2-$\sigma$ confidence levels of some of the free parameters in Eq.~\eqref{eq: 28} (the rest of the parameters not shown here have been marginalized), while imposing Planck 2018 priors~\cite{Planck:2018vyg} on the cosmological parameters. Blue ellipses correspond to  HERA-only forecasts, while the green ellipses account for information coming from CMB-S4 as well. Results are shown for the moderate foreground scenario, although they barely change when pessimistic foreground scenario is considered.}
\label{Fig: figure_6}
\end{figure*}

In the Fisher formalism, the covariance matrix is given by the inverse of the Fisher matrix~\cite{Jungman:1995bz, Mason:2022obt},
\begin{equation}\label{eq: 27}
F_{\alpha,\beta}^\mathrm{21cm}=\sum_{k,z}\frac{\partial\Delta_{21}^2\left(k,z\right)}{\partial\alpha}\frac{\partial\Delta_{21}^2\left(k,z\right)}{\partial\beta}\frac{1}{\left[\delta\Delta_{21}^2\left(k,z\right)\right]^2}.
\end{equation}
Here, $\alpha$ and $\beta$ denote the free parameters that we vary. We vary both cosmological parameters and astrophysical parameters\footnote{Although some of the astrophysical parameters in {\tt 21cmFAST} are defined logarithmically (e.g.\ $L_X^\mathrm{(II)}=40.5$), in our analysis we make sure we vary the linear parameters (e.g.\ $L_X^\mathrm{(II)}=10^{40.5}$). In Fig.~\ref{Fig: figure_6}, when we present the confidence level ellipses of $\log_{10}L_X$, we apply the appropriate Jacobian transformation.}~\cite{Kern:2017ccn},
\begin{eqnarray}\label{eq: 28}
\nonumber\left(\alpha,\beta\right)&\in&\{h,\Omega_m,\Omega_b, A_s,
\\&&\,\,L_X^\mathrm{(II)},f_*^\mathrm{(II)},f_\mathrm{esc}^\mathrm{(II)},A_\mathrm{LW},A_{v_{cb}},T_\mathrm{vir}^{\mathrm{(II})}\}.
\end{eqnarray}
The varied astrophysical parameters in our analysis are displayed in the second row. $L_X^\mathrm{(II)}$ is the X-ray luminosity (normalized by the SFR, in units of $\mathrm{erg}\,\mathrm{sec}^{-1}\,M_\odot^{-1}\,\mathrm{year}$), $f_*^\mathrm{(II)}$ is the star formation efficiency, $f_\mathrm{esc}^\mathrm{(II)}$ is the escape fraction of ionizing photons, $A_\mathrm{LW}$ ($A_{v_{cb}}$) characterizes the amplitude of the LW ($V_{cb}$) feedback on $M_\mathrm{mol,min}$, and $T_\mathrm{vir}^{\mathrm{(II})}$ is the minimum halo virial temperature. Quantities with a super-script (II) correspond to pop-II stars. We also vary the analogous pop-III parameters, around the same fiducial values as the pop-II ones.

The parameters listed in Eq.~\eqref{eq: 28} are not the only parameters that control the 21-cm signal and the CMB. There are many more astrophysical parameters, like the mean-free path of photons through ionized regions, yet for our purposes of demonstrating the joint-analysis of 21-cm and CMB, the astrophysical parameters in Eq.~\eqref{eq: 28} suffice. Another subtlety in our analysis is that we fix $\tau_\mathrm{re}$, although its value can be inferred from the reionization model~\cite{Shmueli:2023box}. A consistent treatment to incorporate $\tau_\mathrm{re}$ in {\tt21cmFirstCLASS} would require computing it from the output of {\tt21cmFAST} and feed its value to {\tt CLASS}. We leave this kind of analysis for future work.

The quantity $\delta\Delta_{21}^2\left(k,z\right)$ that appears in the denominator of Eq.~\eqref{eq: 27} denotes the noise of the experiment, in our case, HERA. We simulate HERA's design sensitivity noise with {\tt 21cmSense}\footnote{\href{https://github.com/steven-murray/21cmSense}{github.com/steven-murray/21cmSense}}~\cite{Pober:2012zz, Pober:2013jna}. In its final stage, HERA will have in its core 331 antennae, each of which has a diameter of $14\,\mathrm{m}$, arranged in a hexagonal array with 11 antennae at its base. We assume the frequency range of HERA will span between $50\,\mathrm{MHz}$ ($z=27.4$) and $225\,\mathrm{MHz}$ ($z=5.3$) with a bandwidth of $8\,\mathrm{MHz}$. This gives a total number of 22 different frequency bands, but in our analysis, to be conservative, we discard the bands below $z\!=\!6$ as in that regime the exact reionization details are highly uncertain, leaving us with 19 redshift bins in total. In each frequency band we assume there are 82 channels, corresponding to 1024 channels over $100\,\mathrm{MHz}$ bandwidth~\cite{DeBoer:2016tnn}. In addition, we assume HERA operates for six hours per night during 540 days in total, the receiver temperature is $T_\mathrm{rec}=100\,\mathrm{K}$ and the sky temperature follows $T_\mathrm{sky}\left(\nu\right)=60\,\mathrm{K}\left(\nu/300\,\mathrm{MHz}\right)^{-2.55}$. 

Finally, we consider in our analysis  ``pessimistic", ``moderate" and ``optimistic" foregrounds scenarios. In the moderate (pessimistic) foregrounds scenario, the wedge\footnote{Our analysis here follows the HERA approach of using only data outside the foreground ``wedge”~\cite{Liu:2019awk, Morales:2012kf}, thus effectively avoiding foregrounds instead of trying to remove them from the data.} is assumed to extend to $\Delta k_{||}=0.1h\,\mathrm{Mpc}^{-1}$ beyond the horizon wedge limit, and all baselines are added coherently (incoherently), while in the optimistic foregrounds scenario the boundary of the foreground wedge is set by the FWHM of the primary beam of HERA and there is no contamination beyond this boundary.

As motivated above, to break degeneracies in the 21-cm signal, we  consider future measurements from CMB-S4~\cite{CMB-S4:2016ple}. We follow Refs.~\cite{Wu:2014hta, Munoz:2016owz, Adi:2020qqf, Shmueli:2023box} and evaluate the Fisher matrix associated with CMB-S4 measurements via
\begin{equation}\label{eq: 29}
F_{\alpha,\beta}^\mathrm{CMB}=\sum_{\ell=30}^{3000}\frac{2\ell+1}{2}f_\mathrm{sky}\mathrm{Tr}\left[C_\mathrm{\ell}^{-1}\frac{\partial C_\ell}{\partial\alpha}C_\mathrm{\ell}^{-1}\frac{\partial C_\ell}{\partial\beta}\right],
\end{equation}
where $f_\mathrm{sky}=40\%$ is the sky-fraction coverage and the matrices $C_\ell\left(\nu\right)$ are (neglecting the lensing contribution)
\begin{equation}\label{eq: 30}
C_\ell\left(\nu\right)=\left[\begin{matrix}
\tilde C_\ell^\mathrm{TT}\left(\nu\right) & C_\ell^\mathrm{TE}\left(\nu\right) \\
C_\ell^\mathrm{TE}\left(\nu\right) & \tilde C_\ell^\mathrm{EE}\left(\nu\right)
\end{matrix}
\right].
\end{equation}
Here, tilde-less quantities are the noise-free CMB anisotropies power spectrum that we take from {\tt CLASS}, while tilde-full quantities include the CMB-S4 noise contribution, $\tilde C_\ell^\mathrm{XX}=C_\ell^\mathrm{XX}+N_\ell^\mathrm{XX}$. The noise power spectra are given by
\begin{equation}\label{eq: 31}
N_\ell^\mathrm{TT}\left(\nu\right)=\Delta_{T}^2\left(\nu\right)\,\mathrm{e}^{\ell\left(\ell+1\right)\sigma_b^2\left(\nu\right)},\quad N_\ell^\mathrm{EE}\left(\nu\right)=2\times N_\ell^\mathrm{TT}\left(\nu\right),
\end{equation}
where $\Delta_T\left(\nu\right)$ is the temperature sensitivity and $\sigma_b\left(\nu\right)=\theta_\mathrm{FWHM}\left(\nu\right)/\sqrt{8\ln 2}$, with the full-width-half-maximum $\theta_\mathrm{FWHM}$ given in radians. We consider a single frequency channel, centered at $\nu=145,\,\mathrm{GHz}$ with  $\Delta_T=1.5\,\mathrm{\mu K\cdot arcmin}$ and $\theta_\mathrm{FWHM}=1.4\,\mathrm{arcmin}$.
Finally, we add the HERA and CMB-S4 Fisher matrices,
\begin{equation}\label{eq: 32}
F_{\alpha,\beta}^{\mathrm{tot}}=F_{\alpha,\beta}^{\mathrm{21cm}}+F_{\alpha,\beta}^{\mathrm{CMB}}.
\end{equation}

\subsection{Forecasts}\label{sec: Forecasts}

Armed with our Fisher formalism, we now vary the free parameters of Eq.~\eqref{eq: 28}, while imposing Planck 2018 priors~\cite{Planck:2018vyg} on the cosmological parameters. In our analysis, we only impose priors on $h$, $\Omega_m$, $\Omega_b$ and $A_s$. Fig.~\ref{Fig: figure_6} shows our results. As expected, adding the CMB-S4 information helps in mitigating all the degeneracies between the different parameters, especially in the cosmological parameters. Because the CMB anisotropies depend only on the cosmological parameters, and the cosmological parameters are not strongly degenerate with the free astrophysical parameters in our analysis, including the information of the CMB power spectra does not help considerably in alleviating degeneracies in the astrophysical parameters. Unlike the cosmological parameters, the well-known  degeneracy between $f_*^\mathrm{(II)}$ and $f_\mathrm{esc}^\mathrm{(II)}$ is evident~\cite{Mason:2022obt, Park:2018ljd, Qin:2021gkn}. These parameters exhibit a negative correlation as the ionization efficiency is proportional to the product of $f_*^\mathrm{(II)}$ and $f_\mathrm{esc}^\mathrm{(II)}$. Similarly, as the X-ray heating efficiency is proportional to the product of $f_*^\mathrm{(II)}$ and $L_X^\mathrm{(II)}$, there is a negative correlation between these parameters as well. Note that the degeneracy of $f_\mathrm{esc}^\mathrm{(II)}$ and $L_X^\mathrm{(II)}$ with $f_*^\mathrm{(II)}$ is not complete because the latter also determines the efficiency of the Ly$\alpha$ flux.

\begin{figure}
\includegraphics[width=\columnwidth]{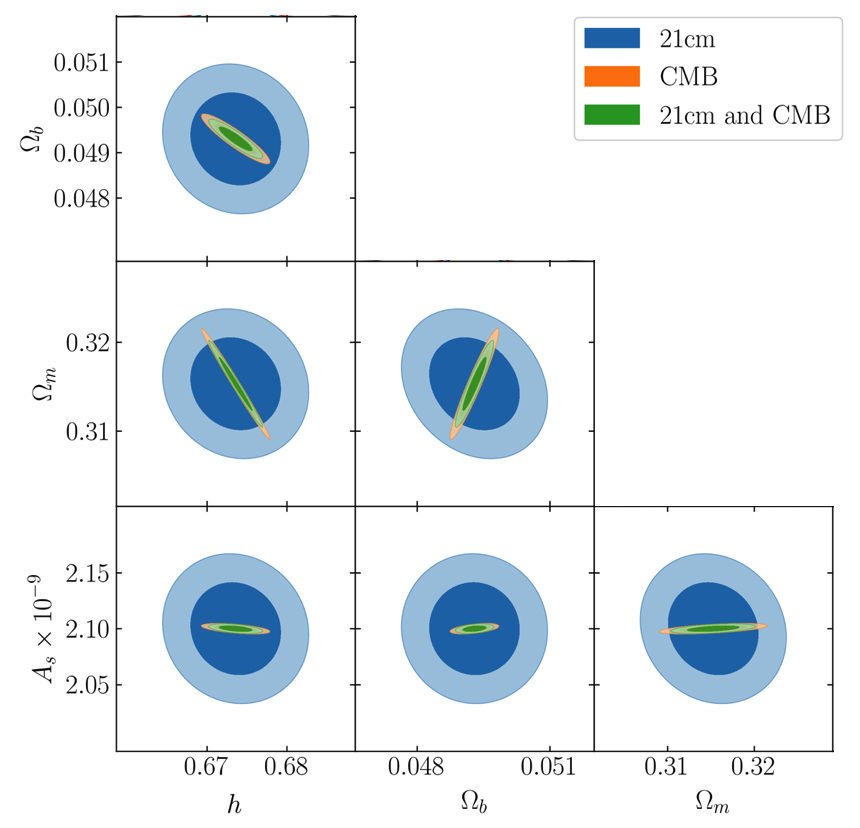}
\vspace{-0.25in}
\caption{Forecasts of 1-$\sigma$ and 2-$\sigma$ confidence levels of the free cosmological parameters in Eq.~\eqref{eq: 28}, while imposing Planck 2018 priors~\cite{Planck:2018vyg} on the cosmological parameters. Blue (orange) ellipses correspond to forecasts when only information from HERA (CMB-S4) is considered, while the green ellipses account for information coming from both HERA and CMB-S4. All the astrophysical parameters have been marginalized (fixed) in the calculation of HERA (CMB-S4) Fisher matrix. For HERA, results are shown for the moderate foreground scenario, although they barely change when the pessimistic foreground scenario is considered.}
\label{Fig: figure_7}
\end{figure}

Not unexpectedly, for $\Lambda$CDM, CMB-S4 will have more constraining power than HERA, as indicated by Fig.~\ref{Fig: figure_7}. Here, all the astrophysical parameters were marginalized (fixed) when only the information of HERA (CMB- S4) was considered. As we shall see in the next section, this statement can be different for beyond $\Lambda$CDM cosmologies, and the 21-cm data can play a more dominant role.

\section{Scattering dark matter}\label{sec: Scattering dark matter}

To demonstrate the potential of {\tt 21cmFirstCLASS} in studying non-linear models beyond the standard model, we now consider SDM. In this model, a fraction $f_\chi$ of the dark matter consists of particles of mass $m_\chi$ that interact directly in a non-gravitational manner with baryons. In this paper, we focus on $f_\chi\!=\!100\%$ and $m_\chi\!=\!1\,\mathrm{MeV}$, although these parameters can be varied in our code (we vary them in Ref.~\cite{FlitterSDMForecasts}). The cross section for the baryons-SDM interaction is parameterized by $\sigma=\sigma_n\left(v/c\right)^{n}$, where $v$ is the relative velocity between the interacting baryon and SDM particles. We also fix $n\!=\!-4$ to correspond to a Coulomb-type interaction (we will relax this assumption in Ref.~\cite{FlitterSDMForecasts}) and thus $\sigma_{-4}$ is the only free parameter in the model we are considering.

There are two consequences to the direct interaction between the baryons and the SDM: (1) they transfer energy, thereby the cold SDM is able to cool the hotter baryonic gas, while the baryons heat the SDM and increase its temperature $T_\chi$, and (2) the bulk relative velocity $V_{\chi b}$ between the two fluids is decreased via a drag force that they apply on each other. The former effect made the SDM a very popular dark matter candidate after the EDGES collaboration announced they measured a minimum value of $\bar T_{21}=-500^{+200}_{-500}\,\mathrm{mK}$ (at $99\%$ confidence level)~\cite{Bowman:2018yin}, which is $3.8\sigma$ below $\Lambda$CDM expectation. More recent results from the SARAS-3 experiment~\cite{Singh:2021mxo} do not reproduce the detection, however.

\subsection{Evolution equations}

Below we write the differential equations that have to be solved in the SDM model. These equations were originally derived in Ref.~\cite{Munoz:2015bca} and appeared since then in many works in the literature. We use a slightly different notation which will be useful for the derivation of the {\it DM-TCA} equations (see Appendix~\ref{sec: Dark matter tight coupling approximation}). The SDM interaction modifies the evolution equation for $T_k$, Eq.~\eqref{eq: 3}, which now reads (note that from this point on we will mostly denote the gas temperature with $T_b$ in order to have symmetrical expressions for the baryons and SDM)
\begin{equation}\label{eq: 11}
\frac{dT_b}{dz}=\frac{dt}{dz}\left[-2HT_b+\Gamma_C\left(T_\gamma-T_b\right)+\frac{2\dot Q_b}{3k_B}+\left.\frac{dT_b}{dt}\right|_\mathrm{ext}\right],
\end{equation}
and a similar equation for the SDM temperature exists,
\begin{equation}\label{eq: 12}
\frac{dT_\chi}{dz}=\frac{dt}{dz}\left[-2HT_\chi+\frac{2\dot Q_\chi}{3k_B}+\left.\frac{dT_\chi}{dt}\right|_\mathrm{ext}\right],
\end{equation}
where
\begin{equation}\label{eq: 13}
\left.\frac{dT_\chi}{dt}\right|_\mathrm{ext}=\frac{2}{3}\frac{T_\chi}{1+\delta_\chi}\frac{d\delta_\chi}{dt},
\end{equation}
and $\delta_\chi\equiv\delta\rho_\chi/\bar\rho_\chi$ is the SDM density contrast. To solve for $T_b$ and $T_\chi$, we need a third differential equation, for the evolution of the bulk relative velocity between the fluids $V_{\chi b}$,
\begin{eqnarray}\label{eq: 14}
\nonumber\frac{dV_{\chi b}}{dz}&=&\frac{dt}{dz}\left[-HV_{\chi b}-D\left(V_{\chi b}\right)\right]
\\&=&\frac{dt}{dz}\left[-HV_{\chi b}-\sum_t D_t\left(V_{\chi b}\right)\right],
\end{eqnarray}
where $D\left(V_{\chi b}\right)$ is the mutual drag force that acts on the baryons and SDM fluids. It is the sum of all the drag forces that arise from the interaction of an SDM particle with a standard model target particle of type $t$,
\begin{equation}\label{eq: 15}
D_t\left(V_{\chi b}\right)=\frac{\rho_\mathrm{tot}\sigma_{-4}c^4}{\rho_bu_{\chi t}^2}\frac{\rho_tF\left(r_t\right)}{m_t+m_\chi}.
\end{equation}
Here, $\rho_b$ ($\rho_\chi=f_\chi\rho_c$) is the baryons (SDM) energy density, $\rho_\mathrm{tot}=\rho_b+\rho_\chi$ (it is not the total matter energy density if $f_\chi<1$), $m_t$ is the mass of the target particle, and $\rho_t$ is the energy density of the target particles. The function $F\left(r_t\right)$ is
\begin{equation}\label{eq: 16}
F\left(r_t\right)=r_t^{-2}\left[\mathrm{erf}\left(\frac{r_t}{\sqrt{2}}\right)-\sqrt{\frac{2}{\pi}}r_t\mathrm{e}^{-r_t^2/2}\right]\underset{r_t\ll1}{\approx}\sqrt{\frac{2}{9\pi}}r_t,
\end{equation}
where $r_t\equiv V_{\chi b}/u_{\chi t}$ and $u_{\chi t}$ is the thermal velocity,
\begin{equation}\label{eq: 17}
u_{\chi t}\equiv\sqrt{\frac{k_BT_b}{m_t}+\frac{k_BT_\chi}{m_\chi}}.
\end{equation}
The cooling/heating rates $\dot Q_b$ and $\dot Q_\chi$ that appear in Eqs.~\eqref{eq: 11} and \eqref{eq: 12} are given by
\begin{eqnarray}\label{eq: 18} 
\nonumber\dot Q_b&=&\frac{3}{2}\Gamma_{\chi b}k_B\left(T_\chi-T_b\right)
\\&&+\frac{\rho_\chi}{\rho_\mathrm{tot}}V_{\chi b}\sum_t\frac{m_\chi m_b}{m_\chi+m_t}D_t\left(V_{\chi b}\right)
\end{eqnarray}
\begin{eqnarray}\label{eq: 19}
\nonumber\dot Q_\chi&=&\frac{3}{2}\frac{n_b}{n_\chi}\Gamma_{\chi b}k_B\left(T_b-T_\chi\right)
\\&&+\frac{\rho_b}{\rho_\mathrm{tot}}V_{\chi b}\sum_t\frac{m_\chi m_t}{m_\chi+m_t}D_t\left(V_{\chi b}\right),
\end{eqnarray}
where $n_b=\rho_b/m_b$ ($n_\chi=\rho_\chi/m_\chi$) is the baryons (SDM) number-density. Finally, the energy transfer rate $\Gamma_{\chi b}$ is
\begin{equation}\label{eq: 20}
\Gamma_{\chi b}=\sqrt{\frac{2}{\pi}}\frac{2\sigma_{-4}c^4\rho_\chi}{3n_b}\sum_t\frac{\rho_t\mathrm{e}^{-r_t^2/2}}{\left(m_t+m_\chi\right)^2u_{\chi t}^3}.
\end{equation}

Two SDM models are typically considered in the literature\footnote{There are also models in which the SDM interacts with either protons or electrons, but not both~\cite{Ali-Haimoud:2015pwa, Gluscevic:2017ywp, Boddy:2018kfv, Maamari:2020aqz, Nguyen:2021cnb, Buen-Abad:2021mvc, Rogers:2021byl}, and there are models in which the SDM directly interacts with CDM~\cite{Liu:2019knx, Barkana:2022hko}.}. The first one considers millicharged DM~\cite{Driskell:2022pax, McDermott:2010pa, Kovetz:2018zan, Munoz:2018pzp, Berlin:2018sjs, Barkana:2018qrx, Slatyer:2018aqg, Liu:2018uzy, Munoz:2018jwq}, in which the target particles are \emph{free} electrons and protons, $n_{t1}=n_{t2}=n_e$, $m_{t1}=m_e$, $m_{t2}=m_p$. Because the number-density of the target particles is proportional to $x_e$, which is very small between recombination and reionization, this model does not generate strong signatures in the 21-cm signal, unless very large cross-sections (that are already ruled out by CMB measuerments) are considered. Instead, we focus on a baryo-philic SDM~\cite{Dvorkin:2013cea, Munoz:2015bca, Boddy:2018wzy, Fialkov:2018xre, Xu:2018efh, Barkana:2018lgd, Short:2022bmm, He:2023dbn, Driskell:2022pax}, in which SDM interacts with \emph{all} standard model particles, i.e.\ $\rho_t=\rho_b$ and $m_t=m_b$, where the mean baryon mass is given by
\begin{equation}\label{eq: 21}
m_b=\frac{m_\mathrm{H}}{\left[1-\left(1-m_\mathrm{H}/m_\mathrm{He}\right)Y_\mathrm{He}\right]\left(1-x_e\right)},
\end{equation}
with $m_\mathrm{H}$ ($m_\mathrm{He}$) the mass of the hydrogen (helium) atom and $Y_\mathrm{He}=\rho_\mathrm{He}/\rho_b\approx0.245$ is the helium mass-fraction.

As in $\Lambda$CDM, we solve Eqs.~\eqref{eq: 11}-\eqref{eq: 14} at each cell using the Euler method, with a step-size of $\Delta z_n=0.1$. At low temperatures, the logarithmic redshift sampling of the standard {\tt 21cmFAST} below $z=35$ is not enough, and we continue to work with $\Delta z_n=0.1$ at the low redshifts regime. Furthermore, we note that attempting to solve Eqs.~\eqref{eq: 11}-\eqref{eq: 14} via the Euler method for large cross-sections results in overshooting of the solution due to the strong coupling between the baryons and the SDM at low redshifts. We therefore had to devise a dedicated method for solving the equations in the strong coupling limit---this is the \emph{DM}-TCA (in contrast with \emph{Compton}-TCA). We elaborate more on that method in Appendix \ref{sec: Dark matter tight coupling approximation}.

\subsection{Initial conditions}\label{subsec: Initial conditions}

To generate the SDM initial conditions for {\tt 21cmFAST} we use a modified version of {\tt CLASS}\footnote{\href{https://github.com/kboddy/class_public/tree/dmeff}{github.com/kboddy/class\_public/tree/dmeff}} in which the SDM fluid variables $\delta_\chi$, $\theta_\chi$, $T_\chi$, are solved simultaneously with the rest of the standard fluid variables of the baryons and CDM (more details on that version can be found in Ref.~\cite{Boddy:2018wzy}). The present total matter density transfer function is then given by $\Omega_m\mathcal T_m=\Omega_c\left[\left(1-f_\chi\right)\mathcal T_c+f_\chi\mathcal T_\chi\right]+\Omega_b\mathcal T_b$. There is a subtlety in the calculation of $\mathcal T_m\left(k,z=0\right)$ that we would like to address. The evolution of $\delta_b$ and $\delta_\chi$ depends on the gas temperature $T_b$ since the momentum exchange rate depends on the thermal velocity $u_{\chi t}$. Even though {\tt CLASS} uses a toy model for the X-ray heating rate $\epsilon_X$ (see Eq.~\eqref{eq: 5}), the resulting transfer function is still correct. The reason for this is that $u_{\chi t}$ competes with $V_{\chi b}$  in the evolution equations of $\delta_b$ and $\delta_\chi$, and since $V_{\chi b}$ becomes very small already at high redshifts (c.f. Fig.~\ref{Fig: figure_8}), $u_{\chi t}$ turns out to have  minimal impact on the low-redshift evolution.

We also extract from {\tt CLASS} the quantity $\mathcal T_{v_{\chi b}}\left(z_\mathrm{rec},k\right)$, the transfer function of the relative velocity between baryons and SDM at recombination, with an equation similar to Eq.~\eqref{eq: 9}. In {\tt 21cmFirstCLASS}, we then generate a $\mathbf V_{\chi b}\left(\mathbf k, z_\mathrm{rec}\right)$ box in Fourier space via~\cite{Munoz:2019rhi}
\begin{equation}\label{eq: 22}
\mathbf V_{\chi b}\left(\mathbf k, z_\mathrm{rec}\right)=i\frac{\mathbf kc}{k}\frac{\mathcal T_{v_{\chi b}}\left(k,z_\mathrm{rec}\right)}{\mathcal T_m\left(k,z=0\right)}\delta_m\left(\mathbf k,z=0\right).
\end{equation}
This yields a $\mathbf V_{\chi b}\left(\mathbf k, z_\mathrm{rec}\right)$ field that is curl-free and completely correlated with $\delta_m\left(\mathbf k,z=0\right)$. In real space, the box of $V_{\chi b}\left(\mathbf x, z_\mathrm{rec}\right)=\left[\mathbf V_{\chi b}\left(\mathbf x, z_\mathrm{rec}\right)\cdot\mathbf V_{\chi b}\left(\mathbf x, z_\mathrm{rec}\right)\right]^{1/2}$ has a Maxwell-Boltzmann distribution with an RMS of
\begin{equation}\label{eq: 23}
\langle V_{\chi b}^2\left(z_\mathrm{rec}\right)\rangle=A_s\int_{k_\mathrm{min}}^{k_\mathrm{max}}\frac{dk}{k}\left(\frac{k}{k_\star}\right)^{n_s-1}\mathcal T^2_{v_{\chi b}}\left(k,z_\mathrm{rec}\right),
\end{equation}
where $k_\mathrm{min}$ ($k_\mathrm{max}$) are determined from the box (cell) size.

Two notes on the above prescription. First, the mean of the $V_{\chi b}\left(\mathbf x, z_\mathrm{rec}\right)$ box is $\langle V_{\chi b}\left(z_\mathrm{rec}\right)\rangle=\sqrt{8/\left(3\pi\right)}\langle V_{\chi b}^2\left(z_\mathrm{rec}\right)\rangle^{1/2}\approx0.92\langle V_{\chi b}^2\left(z_\mathrm{rec}\right)\rangle^{1/2}$. Because of the finite box and cell size this is \emph{not} the true globally-averaged value of $V_{\chi b}$ at recombination. For example, in Fig.~\ref{Fig: figure_8} the initial value of $V_{\chi b}$ in all curves is off by $\sim3\%$. As a consequence, when we plot the mean values of our box at Sec.~\ref{subsec: Results - SDM}, they do not correspond precisely to the true global values. Since in this paper we are mostly interested in the fluctuations of the 21-cm signal, we are not bothered by that nuance. Secondly, the Maxwellianity of $V_{\chi b}$ breaks right after recombination. This is because of the drag term in Eq.~\eqref{eq: 14}, as it renders the differential equation for $V_{\chi b}$  non-linear. Of course, there is no reason to expect that precisely at recombination $V_{\chi b}$ was Maxwellian. In fact, in the derivation of Eqs.~\eqref{eq: 11}-\eqref{eq: 14}, Maxwellianity was assumed throughout. We are therefore being conservative and solve in this work the same equations commonly found in the general SDM literature, despite the inherent inconsistency that this model has. Clearly, the Maxwellianity assumption has to be relaxed, and we leave the study of non-Maxwellianities for future work (see, however, very interesting insights from Refs.~\cite{Ali-Haimoud:2018dvo, Gandhi:2022tmt} on that particular subject).

\subsection{Small temperature corrections}\label{sec: Small temperature corrections}

The direct coupling between SDM and baryons may cause the temperature of the latter to reach very low values, much less than $1\,\mathrm K$ (c.f.\ Fig.~\ref{Fig: figure_9}). This requires modifying some of the key quantities used in {\tt 21cmFAST}.

We take the small temperature correction for the brightness temperature from Ref.~\cite{Barkana:2022hko},
\begin{equation}\label{eq: 24}
T_{21}=\frac{1}{1+z}\left[\frac{\zeta\left(z\right)}{\mathrm{e}^{\zeta\left(z\right)}-1}T_s-T_\gamma\right]\left(1-\mathrm{e}^{-\tau_{21}}\right),
\end{equation}
where $\zeta\left(z\right)=T_\star/T_s\left(z\right)$ and $T_\star=68.2\,\mathrm{mK}$ is the hydrogen hyperfine energy gap (in units of mK). Normally, $T_s\gg T_\star$, and so the new $\zeta$ correction in Eq.~\eqref{eq: 24} approaches 1. When $T_s$ becomes comparable to $T_\star$, the new term becomes important. Yet, because of the following modification, we will see in Sec.~\ref{subsec: Results - SDM} that $T_s$ does not become very small even if $T_b$ does.

The Ly$\alpha$ coupling coefficient $\tilde x_\alpha$ is proportional to the Ly$\alpha$ flux times a correction factor $\tilde S_\alpha$. In the standard {\tt 21cmFAST}, $\tilde S_\alpha$ is evaluated from the fit of Ref.~\cite{Hirata:2005mz}. This fit becomes inadequate at low temperatures (when $T_b\lesssim2\,\mathrm{K}$). We therefore follow Ref.~\cite{Driskell:2022pax} and adopt the wing approximation from Refs.~\cite{Chuzhoy:2005wv, Mittal:2020kjs} to evaluate $\tilde S_\alpha$ (see more details in Appendix~\ref{sec: Small temperature correction for Salpha}).

Another modification that has to be done is in the recombination rate $\alpha_\mathrm{rec}$. In the standard {\tt 21cmFAST}, a fit for the case-A recombination rate is used~\cite{Abel:1996kh}. Again, the validity of this fit breaks at low temperatures. We thus adopt our {\tt HyRec} scheme that was described in Sec.~\ref{sec: Initial conditions} (the SDM does not alter the physics of recombination and so no further modifications in {\tt HyRec} are required).

Finally, we comment that in {\tt 21cmFAST}, the collisional coupling $x_\mathrm{coll}$ is evaluated from tabulated values of $\kappa_{1-0}^\mathrm{iH}$. These are the collision rates of hydrogen atoms with species of type $i$ (in units of $\mathrm{cm^3/sec}$). The tabulated values stop at $T_b=1\,\mathrm{K}$ and the logic of the code is to use $\kappa_{1-0}^\mathrm{iH}\left(T_b=1\,\mathrm{K}\right)$ if $T_b<1\,\mathrm{K}$. Because the extrapolation to lower temperatures is not trivial and is beyond the scope of this paper, we leave it for future work. Having said that, we emphasize that $x_\mathrm{coll}$ is mainly relevant during the dark ages, and thus the forecasts we derive in Sec.~\ref{sec: Combining 21cm and CMB data} (which depend on the physics during cosmic dawn) are insensitive to the exact values of $\kappa_{1-0}^\mathrm{iH}$.

\subsection{Small velocity corrections}\label{sec: Small velocity corrections}

The contribution of pop-III stars comes from halos that are massive enough to host them. In {\tt 21cmFAST}, pop-III stars reside in molecular coolling halos and the aforementioned minimum threshold halo mass is proportional to~\cite{Munoz:2021psm}
\begin{equation}\label{eq: 25}
M_\mathrm{mol,min}\left(\mathbf x,z\right)\propto\left[1+A_{v_{cb}}\frac{V_{cb}\left(\mathbf x,z_\mathrm{rec}\right)}{\langle V^2_{cb}\left(z_\mathrm{rec}\right)\rangle^{1/2}}\right]^{\beta_{v_{cb}}},
\end{equation} 
where $V_{cb}\left(\mathbf x,z_\mathrm{rec}\right)$ is the relative velocity between baryons and CDM at the time of recombination (obtained in a very similar process to the one outlined in Sec.~\ref{subsec: Initial conditions}), and $A_{v_{cb}},\,\beta_{v_{cb}}>0$ are free phenomenological parameters. Note that Eq.~\eqref{eq: 25} is the source for the velocity acoustic oscillations (VAOs)---a standard ruler imprinted on the 21-cm power spectrum at large scales~\cite{Munoz:2019rhi, Munoz:2019fkt, Sarkar:2022mdz}.

In the presence of SDM, there are two dark matter species that hamper pop-III structure formation due to their relative velocities with the baryons---CDM and SDM. We weigh their contributions to $M_\mathrm{mol,min}$ in the following way,
\begin{eqnarray}\label{eq: 26}
\nonumber M_\mathrm{mol,min}\left(\mathbf x,z\right)&\propto&\Bigg\{1+A_{v_{cb}}\Bigg[\left(1-f_\chi\right)\frac{V_{cb}\left(\mathbf x,z_\mathrm{rec}\right)}{\langle V^2_{cb}\left(z_\mathrm{rec}\right)\rangle^{1/2}}
\\&&\hspace{-2mm}+f_\chi\frac{V_{\chi b}\left(\mathbf x,z\right)}{\langle V^2_{cb}\left(z_\mathrm{rec}\right)\rangle^{1/2}}\frac{1+z_\mathrm{rec}}{1+z}\Bigg]\Bigg\}^{\beta_{v_{cb}}}.
\end{eqnarray}
The reason for this modelling is because of the following. If $f_\chi=0$ then Eq.~\eqref{eq: 26} becomes identical to Eq.~\eqref{eq: 25}. If $f_\chi\approx1$, then the second term in Eq.~\eqref{eq: 26} dominates. Note that in the special case of $f_\chi\approx1$ and very small $\sigma_{-4}$, SDM behaves as CDM, $V_{\chi b}\approx V_{cb}\propto\left(1+z\right)$, and Eq.~\eqref{eq: 25} is again restored in that scenario. For cross-sections large enough, $V_{\chi b}\ll V_{cb}$ (c.f. Fig.~\ref{Fig: figure_8}). Thus, in an SDM universe, Eq.~\eqref{eq: 26} implies that $M_\mathrm{mol,min}$ is smaller and hence more pop-III stars can be born, thereby pulling cosmic dawn to higher redshifts. It is worthwhile to note that the free parameters $A_{v_{cb}}$ and $\beta_{v_{cb}}$ were calibrated in~\cite{Munoz:2021psm} to match with hydrodynamical N-body simulations~\cite{Kulkarni:2020ovu,ENZO:2013hhu}. As fitting our model to such simulations is beyond the scope of this work, we adopt the simple model of Eq.~\eqref{eq: 26} with the same fiducial parameters. We defer the exploration of more complicated prescriptions (where, for instance, the weights $f_\chi$ and $1-f_\chi$ include power-laws with free indices) to future work.

The SFRD in {\tt 21cmFAST} depends both on $M_\mathrm{mol,min}$ and on the halo mass function (HMF). The latter is modified by SDM in two ways. First, the matter-density variance $\sigma\left(M\right)$ is reduced because of the suppression in the matter power spectrum~\cite{Driskell:2022pax}. And secondly, the fitting function that is used for the evaluation of the HMF is modified. In this work, the former effect  is already taken into account in our analysis, while the second is not. We use the Sheth-Tormen fitting function~\cite{Sheth:1999mn}, which was calibrated based on CDM N-body simulations. It is not clear how the fitting parameters of the Sheth-Tormen HMF are modified if SDM is considered instead of CDM. We leave the exploration of this subtlety\footnote{We thank Mihir Kulkarni for drawing our attention to this assumption in our analysis.} for future work.

\subsection{Results - SDM}\label{subsec: Results - SDM}

\begin{figure}
\includegraphics[width=\columnwidth]{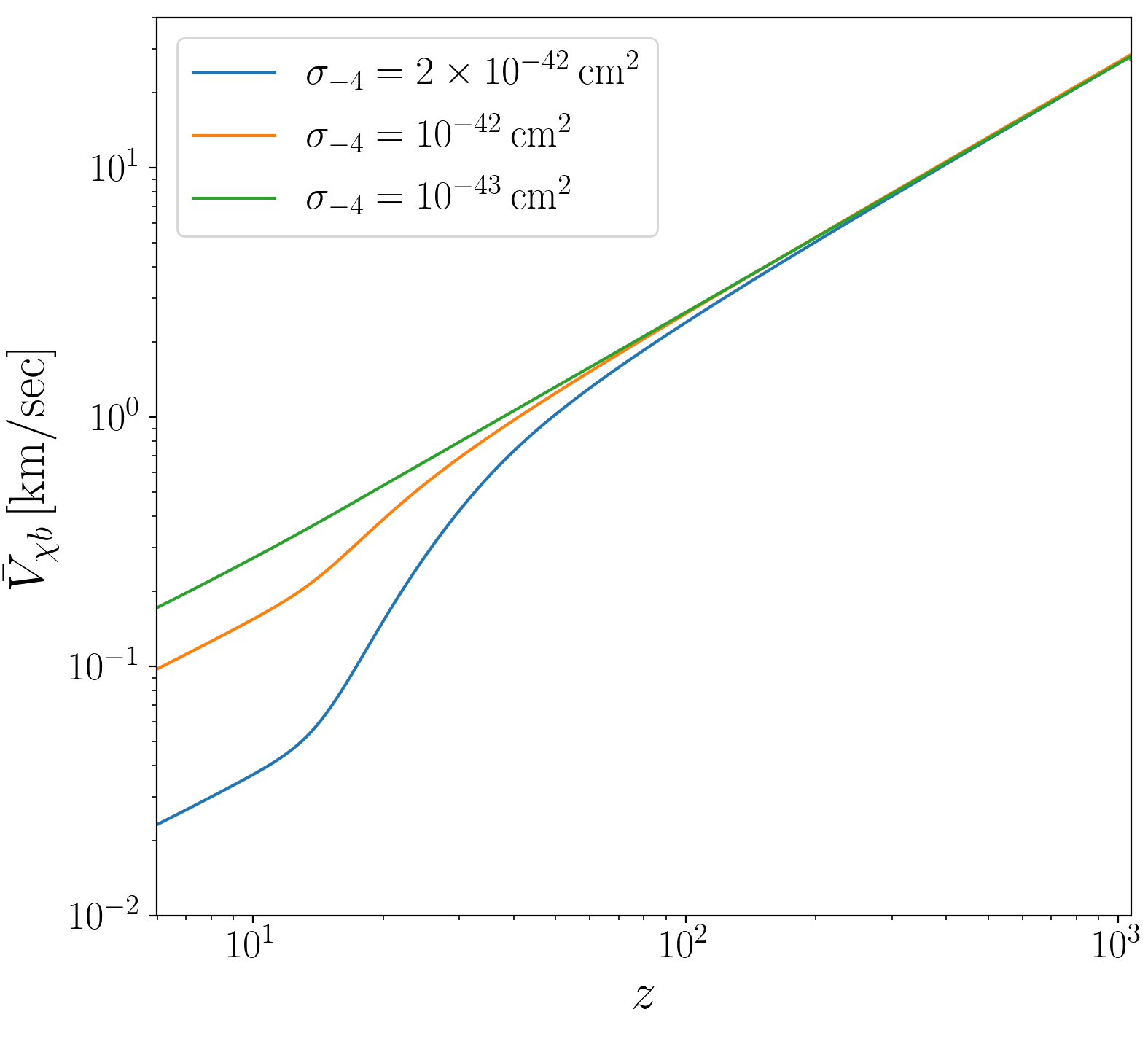}
\caption{Evolution of the ``global" $V_{\chi b}$ (see caveat below Eq.~\eqref{eq: 23}) for three different cross-sections. In this figure and in the upcoming figures we fix $m_\chi=1\,\mathrm{MeV}$ and $f_\chi=100\%$.}
\label{Fig: figure_8}
\end{figure}

In what follows we will consider three case studies where $\sigma_{-4}$ is equal to  $2\times10^{-42}\,\mathrm{cm^2}$, $10^{-42}\,\mathrm{cm^2}$ or $10^{-43}\,\mathrm{cm^2}$. The impact of these cross-sections on the evolution of the baryons and SDM fluids is most clearly seen in Fig.~\ref{Fig: figure_8} where we plot the ``global" $V_{\chi b}$ (see caveat below Eq.~\eqref{eq: 23}). The green curve that corresponds to $\sigma_{-4}=10^{-43}\,\mathrm{cm^2}$ can be considered as ``almost $\Lambda$CDM" because $V_{\chi b}\propto\left(1+z\right)$, which is indeed the evolution of $V_{cb}$ when there is no drag term in Eq.~\eqref{eq: 14}. In contrast, the blue curve of $\sigma_{-4}=2\times10^{-42}\,\mathrm{cm^2}$ decays very quickly; initially the Hubble term dominates, then at $z\sim100$ the drag term wins, and finally at $z\sim15$ the Hubble term dominates again once $V_{\chi b}$ is small enough. The case of $\sigma_{-4}=10^{-42}\,\mathrm{cm^2}$ (orange curve) exhibits a similar decay, although milder. 

\begin{figure}
\includegraphics[width=\columnwidth]{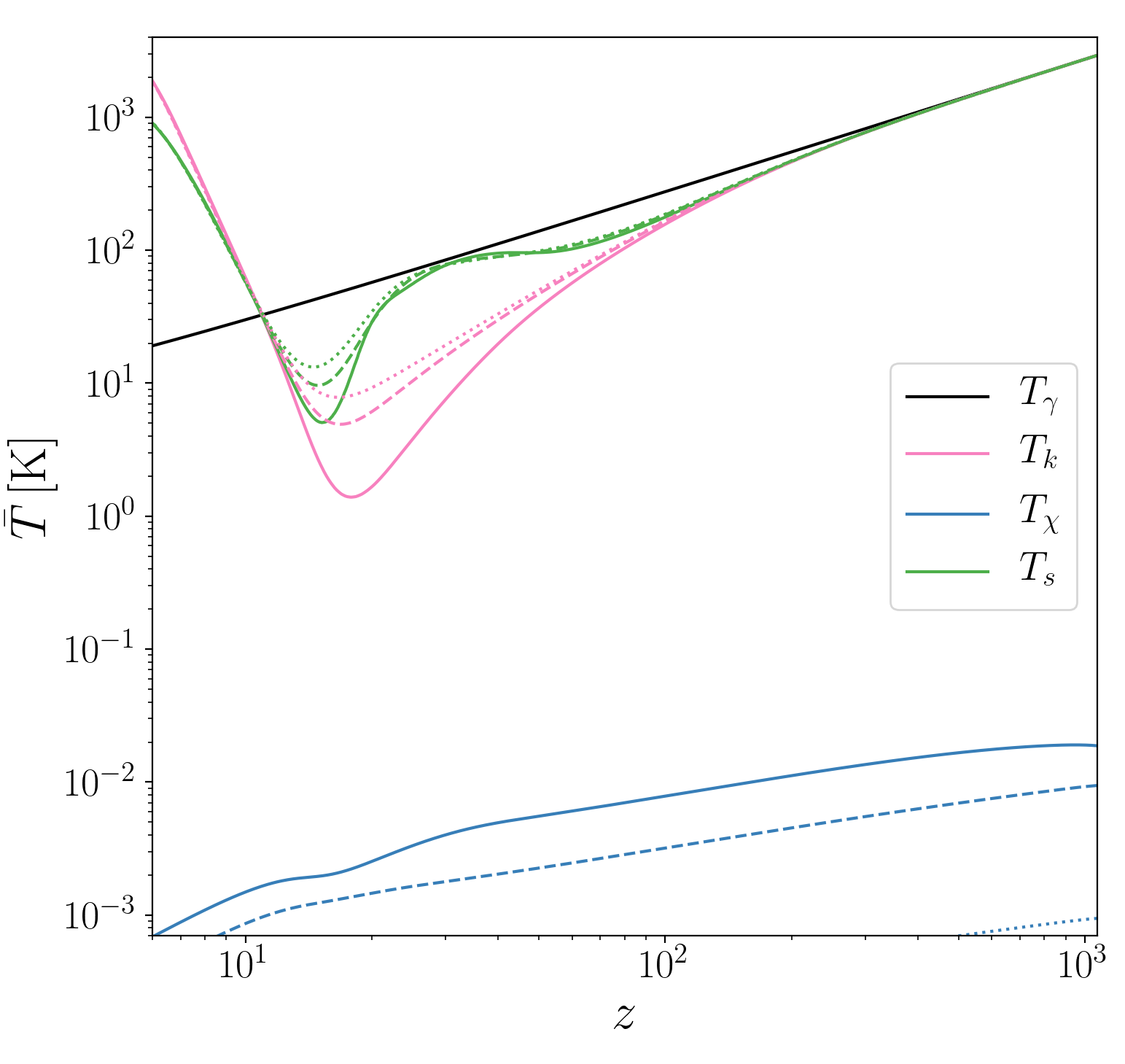}
\caption{Evolution of the global gas kinetic temperature, SDM kinetic temperature and the spin temperature, for three different cross-sections. Solid, dashed and dotted lines correspond to $\sigma_{-4}=2\times10^{-42},\,10^{-42},\,10^{-43}\,\mathrm{cm^2}$, respectively. We note that the $T_s$ solid green curve is most likely too high between $20\lesssim z\lesssim 30$, see text for further details.}
\label{Fig: figure_9}
\end{figure}

Next, we consider the evolution of $T_b$ and $T_\chi$ as it appears in Fig.~\ref{Fig: figure_9}. Let us focus first on the solid curves of $\sigma_{-4}=2\times10^{-42}\,\mathrm{cm^2}$ where the new-physics is most extreme. As expected, the rapid interactions between the baryons and the cold SDM cools down the former considerably. Once stars have been formed, their radiated X-rays heat up the gas, as in $\Lambda$CDM. Note that the turning-point of the pink solid curve appears before the other pink curves, this is because a very cold baryonic gas reacts to the slightest source of heating. In fact, without X-rays, the baryons would have been tightly coupled to the SDM at $z\sim17$ because the interaction rate increases as $V_{\chi b}$ decreases, and $V_{\chi b}$ already approaches zero at low redshifts. As for $T_\chi$, we can see that the Hubble cooling in Eq.~\eqref{eq: 12} mostly dominates w.r.t\ the $\dot Q_\chi$ heating term. Unlike the baryons, which undergo a lot of SDM scattering, the SDM particles barely feel the baryons. This is because $\rho_\chi$ is comparable to $\rho_b$ for $f_\chi=100\%$. However, their number-densities are not; $m_b\approx1\,\mathrm{GeV}\gg m_\chi$ in the model that we are considering and thus $n_\chi\gg n_b$, namely the SDM particles vastly outnumber the baryons. Nevertheless, the SDM is not completely oblivious to the presence of the baryons and it begins to heat-up at $z\sim15$ once the temperature difference becomes large enough. Then, at $z\sim10$ the Hubble cooling wins again and the SDM is further cooled-down. All the physics discussed above applies as well to the dashed and dotted curves in Fig.~\ref{Fig: figure_9}, although to a much lesser extent.

Fig.~\ref{Fig: figure_9} also presents the evolution of the spin temperature. Let us begin the discussion this time with the dashed and dotted curves that correspond to $\sigma_{-4}=10^{-42}\,\mathrm{cm^2}$ and $\sigma_{-4}=10^{-43}\,\mathrm{cm^2}$, respectively. It appears that the WF coupling is stronger for the dashed curve and thus the cosmic dawn allegedly arrives earlier when the cross-section is larger. In the SDM model there are many factors that affect the onset of cosmic dawn. For example, as Ref.~\cite{Driskell:2022pax} pointed out, the matter power spectrum is suppressed on small scales due to the presence of the SDM. This fact contributes to the delaying of cosmic dawn (in a similar mechanism as in FDM~\cite{Sarkar:2022dvl, Flitter:2022pzf}). However, there are other competing effects. First, since $T_\alpha\approx T_k$, smaller $T_k$ tends to drive $T_s$ to smaller values (note however that this effect has nothing to do with the onset of cosmic dawn). Secondly, the lower $V_{\chi b}$ values imply a smaller $M_\mathrm{mol,min}$ (c.f. Eq.~\eqref{eq: 26}), which means that smaller halos (that are much more abundant) can form stars more easily. On the other hand, there are two more effects that tend to weaken the coupling of $T_k$ to $T_s$ for larger cross-sections: (1) Smaller $T_k$ implies smaller $\tilde S_\alpha$ (see Appendix \ref{sec: Small temperature correction for Salpha}), and (2) smaller $M_\mathrm{mol,min}$ leads to a stronger LW radiation that impedes stars formation (although the LW feedback effect may yield a weaker LW flux, so it is not clear a priori if this effect enhances or degrades the coupling of $T_s$ to $T_k$). All in all, we find that for the model that we are considering\footnote{We did witness a weaker WF coupling with much stronger cross-sections, or when we considered $m_\chi=1\,\mathrm{GeV}$.}, $T_s$ is more strongly coupled to $T_k$ when the cross-section is larger. 

As for the solid green curve in Fig.~\ref{Fig: figure_9}, the similar trend continues. For the fiducial parameters we have used, we find that cross-sections larger than $\sigma_{-4}=2\times10^{-42}\,\mathrm{cm^2}$ can result in extremely low temperatures that reach below $T_k=1\,\mathrm{K}$. As was discussed at the end of Sec.~\ref{sec: Small temperature corrections}, {\tt 21cmFAST} interpolates a table of $\kappa_{1-0}^\mathrm{iH}$ that has no entries below $T_k=1\,\mathrm{K}$. Thus, unless the interpolation table of {\tt 21cmFAST} is extended to lower temperatures, larger cross-sections may yield unphysical behaviors at the $T_s$ curve.

\begin{figure}
\includegraphics[width=\columnwidth]{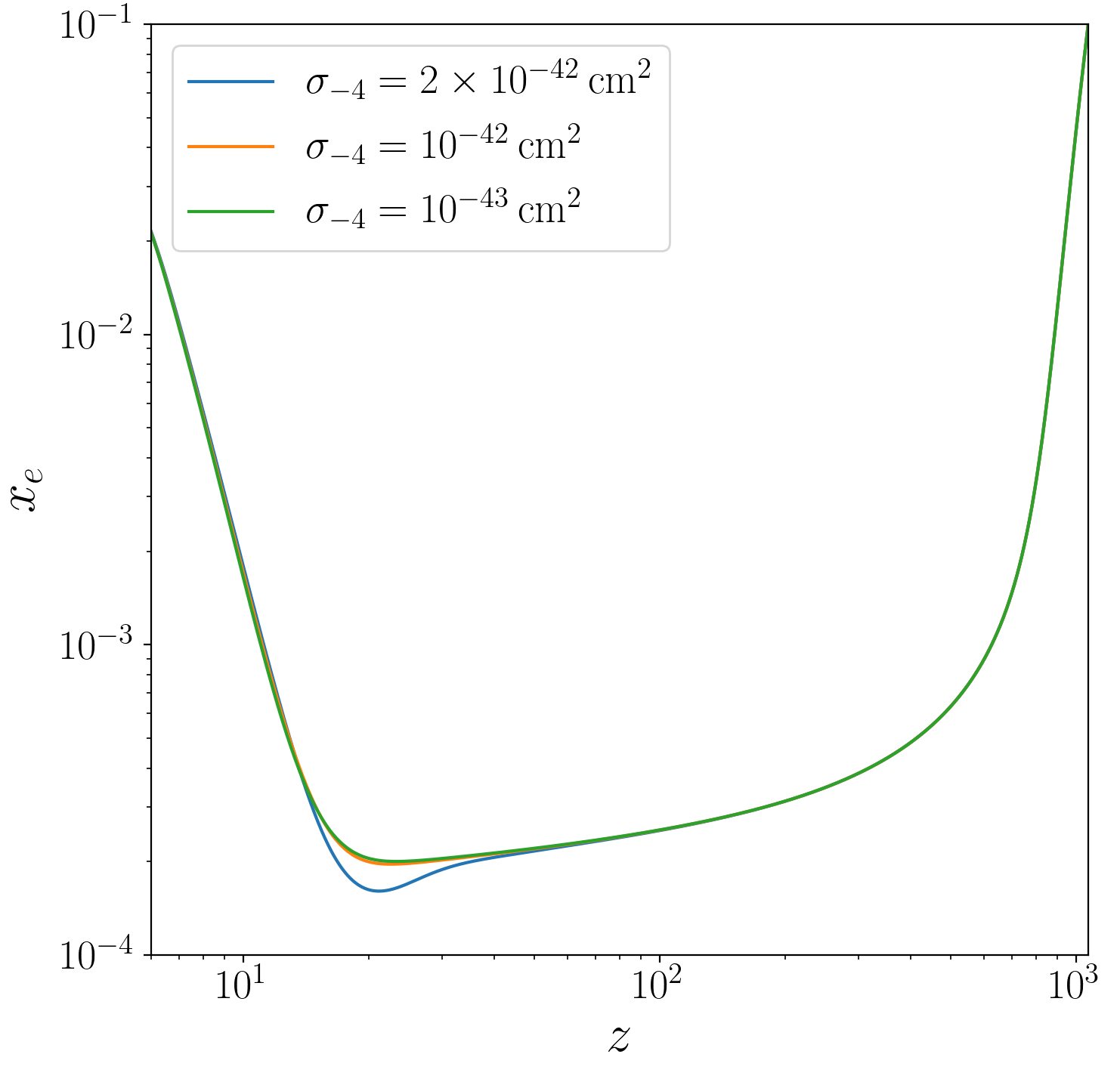}
\caption{Evolution of the global $x_e$ in SDM universe, for three different cross-sections. The green curve of $\sigma_{-4}=10^{-43}\,\mathrm{cm^2}$ is practically indistinguishable from the $\Lambda$CDM curve shown in Fig.~\ref{Fig: figure_3}. The extra drop seen in the blue curve of $\sigma_{-4}=2\times10^{-42}\,\mathrm{cm^2}$, although it can be physically justified, is subject to theoretical uncertainties, see main text for more details.}
\label{Fig: figure_10}
\end{figure}

\begin{figure}
\includegraphics[width=\columnwidth]{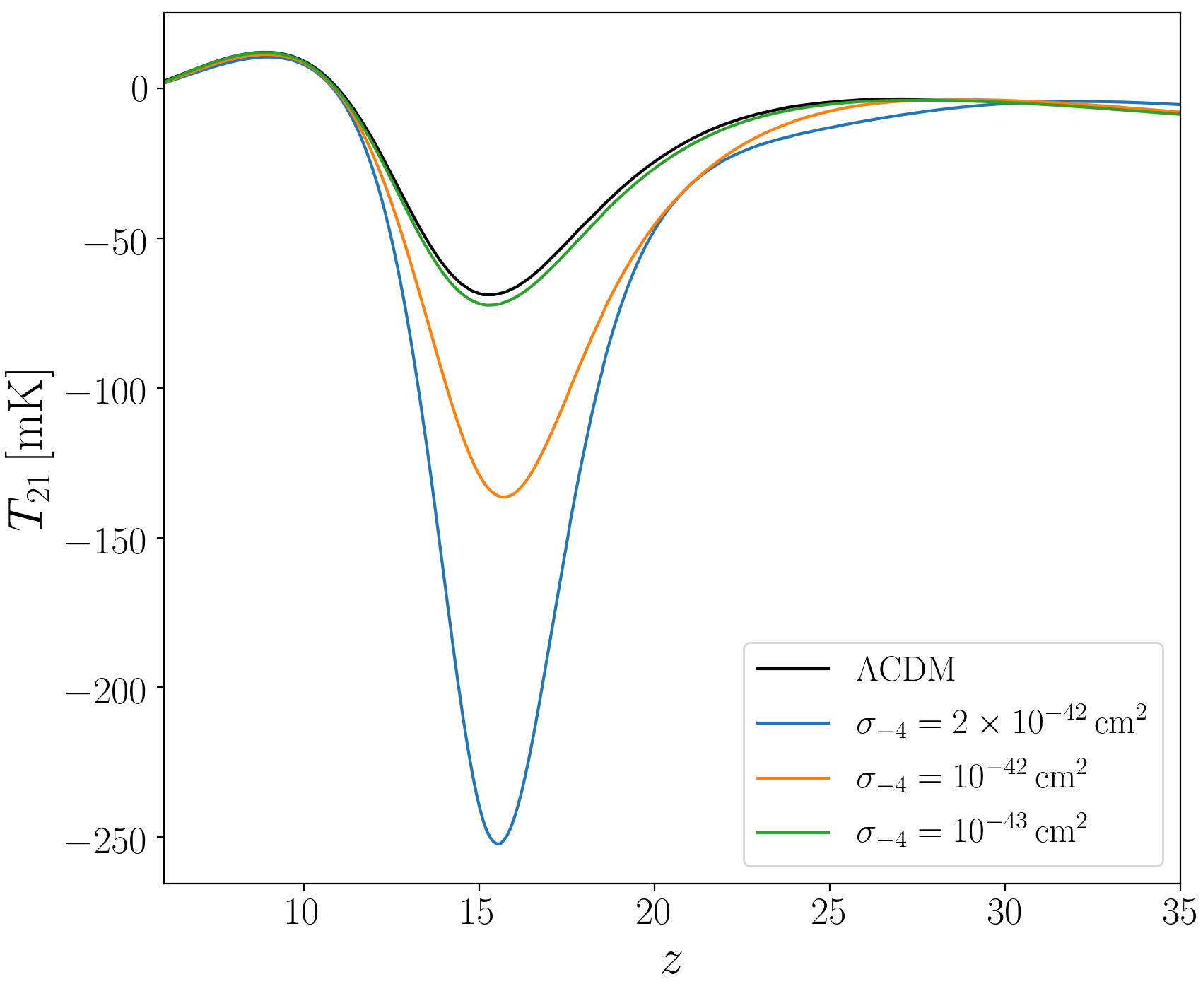}
\caption{The global 21cm signal in SDM universe, for three different cross-sections.}
\label{Fig: figure_11}
\end{figure}

\begin{figure}
\includegraphics[width=0.95\columnwidth]{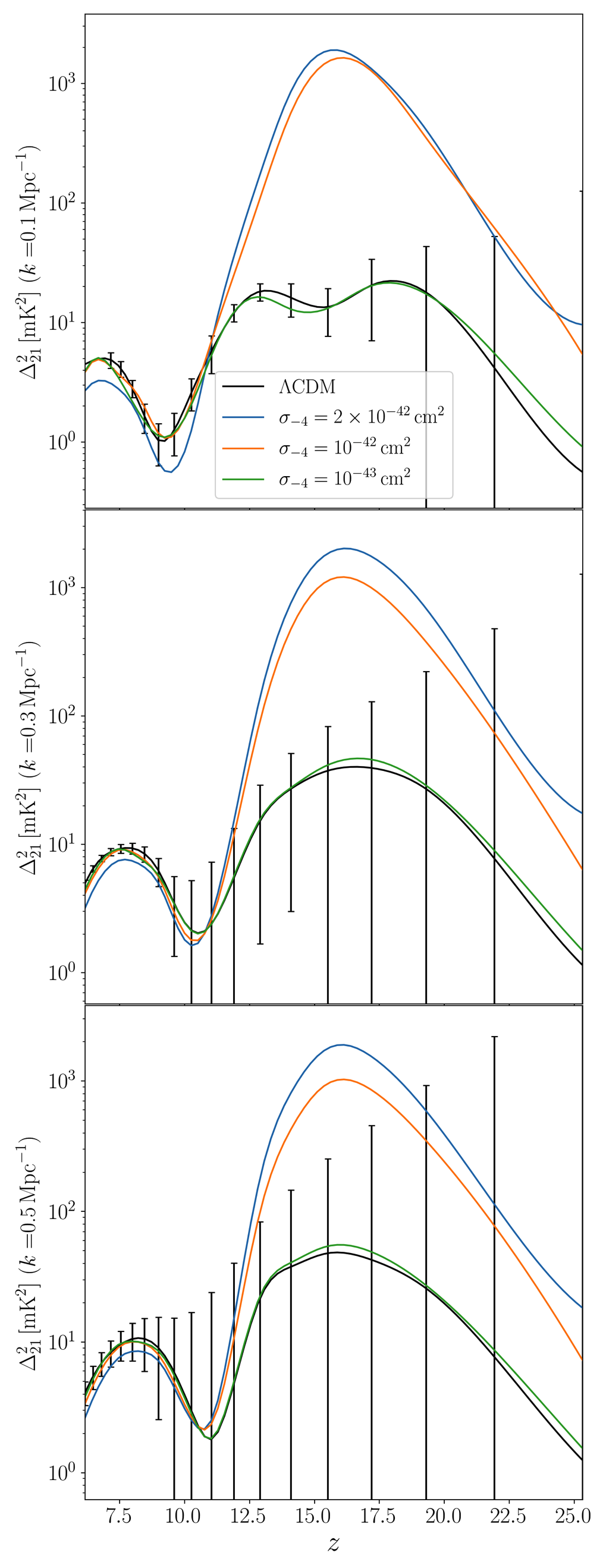}
\caption{The 21cm power spectrum in SDM universe, for three different cross-sections. Here, unlike Fig.~\ref{Fig: figure_5}, we assume moderate foreground scenario for the error bars.}
\label{Fig: figure_12}
\end{figure}

It is also interesting to inspect the evolution of $x_e$ in an SDM universe. For $\sigma_{-4}=10^{-42}\,\mathrm{cm^2}$ and $\sigma_{-4}=10^{-43}\,\mathrm{cm^2}$, Fig.~\ref{Fig: figure_10} shows that SDM barely makes any difference in the evolution of $x_e$ compared to $\Lambda$CDM. However, a surprising feature can be seen when we consider $\sigma_{-4}=2\times10^{-42}\,\mathrm{cm^2}$; at $z\sim40$ we see that $x_e$ departs from the $\Lambda$CDM expectation towards lower values. Normally, in $\Lambda$CDM the temperature of the baryons at this redshift is insufficient to allow an efficient recombination, because their number-density is too low. But for the SDM that we are considering, recombination becomes efficient again at $z\sim40$ because baryons are combined into atoms more easily when the temperature decreases. Without X-ray heating, we find that for this scenario $x_e$ would stabilize on a lower freezout value of $\sim2\times10^{-6}$. Yet, it is important to stress that at low temperatures {\tt HyRec} uses the fit of Ref.~\cite{1991A&A...251..680P} for the recombination rate, but this fit is valid only to $T_k=40\,\mathrm{K}$, so the second drop in $x_e$ shown in Fig.~\ref{Fig: figure_10} should not be taken too seriously.

\begin{figure*}
\includegraphics[width=0.8\textwidth]{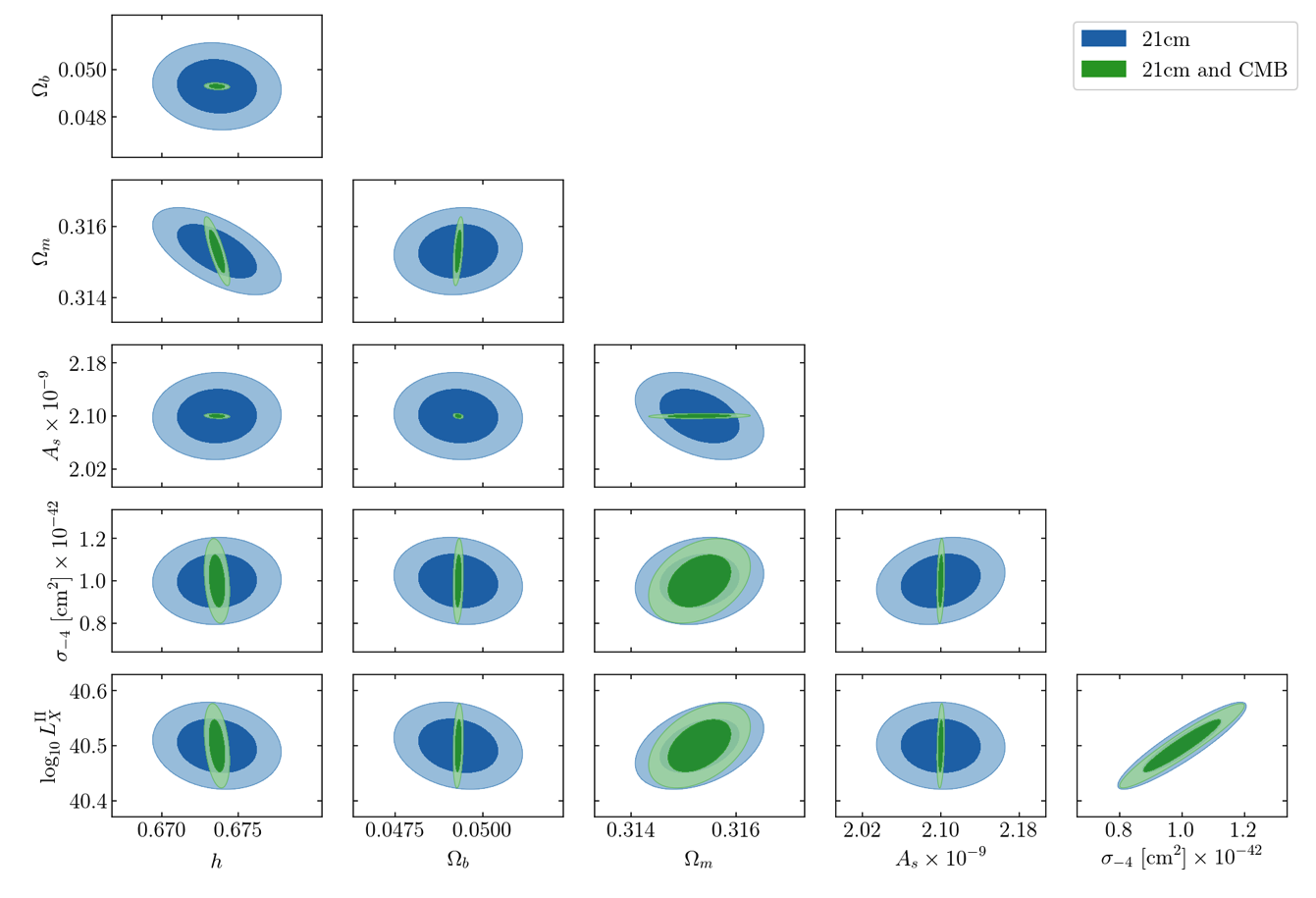}
\caption{Forecasts of 1-$\sigma$ and 2-$\sigma$ confidence levels of some of the free parameters in Eq.~\eqref{eq: 28} and the SDM cross-section $\sigma_{-4}$ (the rest of the parameters not shown here have been marginalized), while imposing Planck 2018 priors~\cite{Planck:2018vyg} on the $\Lambda$CDM cosmological parameters. Blue ellipses correspond to forecasts when only information from HERA is considered, while the green ellipses account for information coming from CMB-S4 as well. Results are shown for moderate foreground scenario, although they barely change when pessimistic foreground scenario is considered.}
\label{Fig: figure_13}
\end{figure*}

The global 21-cm signal, shown in Fig.~\ref{Fig: figure_11}, reflects the same physics previously discussed. Larger cross-sections lead to a deeper absorption signal that begins at higher redshifts, but ends roughly at the same redshift. We show the corresponding 21-cm power spectrum in Fig.~\ref{Fig: figure_12}. It is clearly seen that HERA will be challenged to distinguish between $\Lambda$CDM and SDM of cross-section $\sigma_{-4}=10^{-43}\,\mathrm{cm^2}$. In contrast, it appears that HERA will be able to easily detect SDM with cross-section $\sigma_{-4}=10^{-42}\,\mathrm{cm^2}$ (or higher) but only in the low frequency band that corresponds to $10\lesssim z\lesssim 20$. Two remarks on the blue curve of $\sigma_{-4}=2\times10^{-42}\,\mathrm{cm^2}$: (1) Although its global signal reaches much lower values than the orange curve of $\sigma_{-4}=10^{-42}\,\mathrm{cm^2}$, the amplitude of the 21-cm power spectrum for both cross-sections is of the same order of magnitude. This is most likely because the absorption profile of $\sigma_{-4}=2\times10^{-42}\,\mathrm{cm^2}$ is quite narrow and we calculate the power spectrum from slices of the lightcone box, which unlike the coeval box contains samples from different redshifts along the line-of-sight. (2) The smaller power at low redshifts is due to a shallower emission profile which is caused by the large cooling effect.

\subsection{Forecasts - SDM}\label{sec: Forecasts - SDM}

\begin{figure}
\includegraphics[width=\columnwidth]{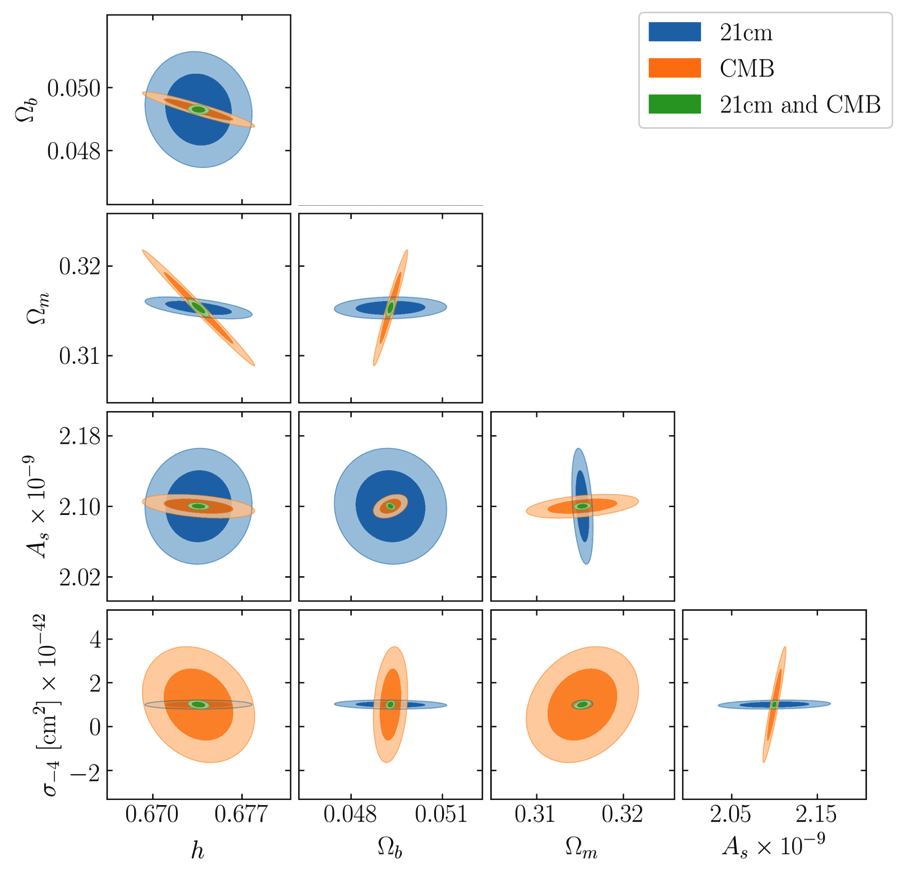}
\caption{Forecasts of 1-$\sigma$ and 2-$\sigma$ confidence levels of the free cosmological parameters in Eq.~\eqref{eq: 28} and the SDM cross-section $\sigma_{-4}$, while imposing Planck 2018 priors~\cite{Planck:2018vyg} on the $\Lambda$CDM cosmological parameters. The assumptions made in this figure are similar to those of Fig.~\ref{Fig: figure_13}, but here we have marginalized over all astrophysical parameters, allowing us to present forecasts for only CMB-S4. Blue (orange) ellipses correspond to forecasts when only information from HERA (CMB-S4) is considered, while the green ellipses account for information coming from both HERA and CMB-S4. All the astrophysical parameters have been marginalized (fixed) in the calculation of the HERA (CMB-S4) Fisher matrix. For HERA, results are shown for the moderate foreground scenario, although they barely change when a pessimistic foreground scenario is considered.}
\label{Fig: figure_14}
\end{figure}

Fig.~\ref{Fig: figure_12} suggests that HERA will not be sensitive to cross-sections below $10^{-43}\,\mathrm{cm^2}$, but cross-sections of the order of $10^{-42}\,\mathrm{cm^2}$ or higher can be probed. Yet, all we did in Fig.~\ref{Fig: figure_12} was to vary the cross-section while keeping other parameters fixed. If we wish to forecast the sensitivity of HERA to SDM, we must vary other cosmological and astrophysical parameters and study their degeneracies, like we did in Sec.~\ref{sec: Combining 21cm and CMB data}. For the following analysis, we focus on the SDM scenario where $\sigma_{-4}=10^{-42}\,\mathrm{cm^2}$. This particular value has not been ruled out by Planck 2018 CMB measurements and it lies beyond the sensitivity range of CMB-S4 by almost an order of magnitude~\cite{Boddy:2018wzy, Xu:2018efh}.

Our forecasts are displayed in Fig.~\ref{Fig: figure_13}. Interestingly, the forecasts for $\Omega_m$ seem to be less affected when combining the information from the two observables. We will see shortly why. Furthermore, we see a strong degeneracy between $\sigma_{-4}$ and $L_X^\mathrm{(II)}$. This feature in our forecasts is not surprising; stronger $\sigma_{-4}$ yields more efficient cooling, while stronger $L_X^\mathrm{(II)}$ yields more efficient heating, thereby any small correlated variation in both of them is almost canceled in the observed brightness temperature. Hence, these two parameters exhibit a positive correlation. Since the CMB anisotropies do not depend on the value of $L_X^\mathrm{(II)}$, their measurement cannot relax this degeneracy.

It is also interesting to compare HERA's performance in detecting SDM with CMB-S4. We make this comparison in Fig.~\ref{Fig: figure_14}. As we saw in Fig.~\ref{Fig: figure_7} when we discussed degeneracies in $\Lambda$CDM, for most cosmological parameters CMB-S4 has a better constraining power than HERA. However, in the  SDM scenario, HERA has the upper hand when it comes to constraining $\Omega_m$ (which is now comprised of SDM, unlike in $\Lambda$CDM) and $\sigma_{-4}$. In particular, for SDM with $\sigma_{-4}=10^{-42}\,\mathrm{cm^2}$, HERA will be able to constrain its value within 2-$\sigma$ confidence level, while CMB-S4 will barely be able to do so within 1-$\sigma$ confidence level. This demonstrates the potential of HERA in detecting new-physics that cannot be probed by CMB-S4.

\section{Conclusions}\label{sec: Conclusions}

In this paper we have introduced our novel pipeline, {\tt 21cmFirstCLASS}, for studying the cosmological 21-cm signal and its anisotropies. It is composed of two codes that are commonly used in the literature---{\tt CLASS} and {\tt 21cmFAST}. Because {\tt CLASS} provides the proper initial conditions for the simulation, as well as the more precise scale-independent growth factor, our code in that sense is more consistent than the standard {\tt 21cmFAST}. Moreover, since our simulation begins from recombination, our calculations naturally capture early temperature and ionization fluctuations, an effect which distorts the 21-cm power spectrum to some extent\footnote{We elaborate more on that subtle point in  Paper II.} (c.f.\ Fig.~\ref{Fig: figure_5}).  To achieve the most precise evolution of the early Universe, we have incorporated in {\tt 21cmFirstCLASS} the state-of-the-art recombination code {\tt HyRec} as an integral part of our calculation.

Unlike {\tt 21cmFAST}, our code is \emph{not} fast. For the box settings we used in this work, starting the simulation at recombination results in a runtime which is $\sim3$ times longer compared to the normal {\tt 21cmFAST} simulation that begins at $z=35$, even though no complicated astrophysics calculations are performed at high redshifts. The runtime ratio becomes even greater when either SDM (which requires more redshift samples below $z=35$) or higher resolution boxes are considered. The source for this longer runtime is the huge amount of redshift samples used in {\tt 21cmFirstCLASS} and the current architecture of {\tt 21cmFAST}; at each redshift iteration the evolution of the box is done at the C-level (where multiple CPUs can facilitate the computation) but at the end of the iteration the box is transferred back to the python-wrapper, where the box can be processed with only a single CPU. We therefore think that changing the {\tt 21cmFAST} architecture such that the C-code will be able to promote the box over more than one redshift iteration may speed-up significantly the calculations of {\tt 21cmFirstCLASS}. Implementing this is beyond the scope of this paper and we defer this necessary modification for future work.

One of the main motivations to begin the simulation from recombination is to study highly non-linear models. As a case study, we focused on SDM, which is one of the most popular candidates of dark matter in the recent literature. This required us using the modified {\tt CLASS} version of Ref.~\cite{Boddy:2018wzy} to get the correct initial conditions. Moreover, besides implementing the SDM differential equations in {\tt 21cmFAST}, we had to make several modifications in the astrophysics part, the most important one is the correction factor $\tilde S_\alpha$ for the WF coupling. As a first thorough study of the effect that SDM has on the 21-cm power spectrum, we limited ourselves to SDM with parameters $f_\chi=100\%$, $m_\chi=1\,\mathrm{MeV}$ and a velocity-dependent cross-section with a power-law of $n=-4$. For very large cross-sections that change the 21-cm signal extremely, our results suffer from an inconsistency at $z\gtrsim20$ due to an approximated modelling of the collisional rates $\kappa_{1-0}^\mathrm{iH}$ at low temperatures. For milder cross sections, our results are consistent at all redshifts.

Focusing on $\sigma_{-4}=10^{-42}\,\mathrm{cm^2}$, which on the one hand has not been ruled out by Planck 2018 measurements, but on the other hand lies beyond the CMB-S4 sensitivity range, we found that HERA in its design sensitivity will be able to easily probe SDM with that cross-section within 2-$\sigma$ confidence level, under the assumption of either moderate or pessimistic foregrounds scenarios, and taking the degeneracies with astrophysical parameters into account. This serves as clear evidence to the very promising potential of HERA and the 21-cm signal in searching for signatures of physics beyond $\Lambda$CDM, provided that state-of-the-art, first-class codes are used.

\begin{acknowledgments}
It is our pleasure to thank Bradley Greig, Julian B. Mu\~noz, Kimberly K. Boddy, Sarah Libanore, Manuel A. Buen-Abad, Hovav Lazare and Gali Shmueli for useful discussions. Furthermore, we would like to thank the anonymous referee for useful comments that improved the quality of the paper. We also acknowledge the efforts of the  {\tt 21cmFAST} and {\tt CLASS} authors to produce state-of-the-art public 21-cm and CMB codes. JF is supported by the Zin fellowship awarded by the BGU Kreitmann School. EDK acknowledges support from an Azrieli faculty fellowship. EDK also acknowledges joint support from the U.S.-Israel Bi-national Science Foundation (BSF,  grant No. 2022743) and the U.S. National Science Foundation (NSF, grant No. 2307354), and support from the ISF-NSFC joint research program (grant No. 3156/23).
\end{acknowledgments}

\appendix
\section{Scale-independent growth factor}\label{sec: Scale-independent growth factor}
Because CDM is collisionless and comprises most of the matter in the Universe, its evolution is nearly scale invariant (especially at high redshifts before baryons have clustered) and thus the growth in its density contrast is given by $\delta_c\left(k,z\right)=D\left(z\right)\delta_c\left(k,z=0\right)$, where $D\left(z\right)$ is the scale-independent growth factor. Using the continuity and Euler equations, together with the Poisson equation, one can show that the differential equation that governs $D\left(z\right)$ is~\cite{baumann_2022} 
\begin{equation}\label{eq: A1}
\ddot D+2H\dot D-4\pi G\bar\rho_mD=0,
\end{equation}
where $G$ is Newton's gravitational constant and over-dots represent derivatives with respect to the cosmological time $t$.

\begin{figure}
\includegraphics[width=\columnwidth]{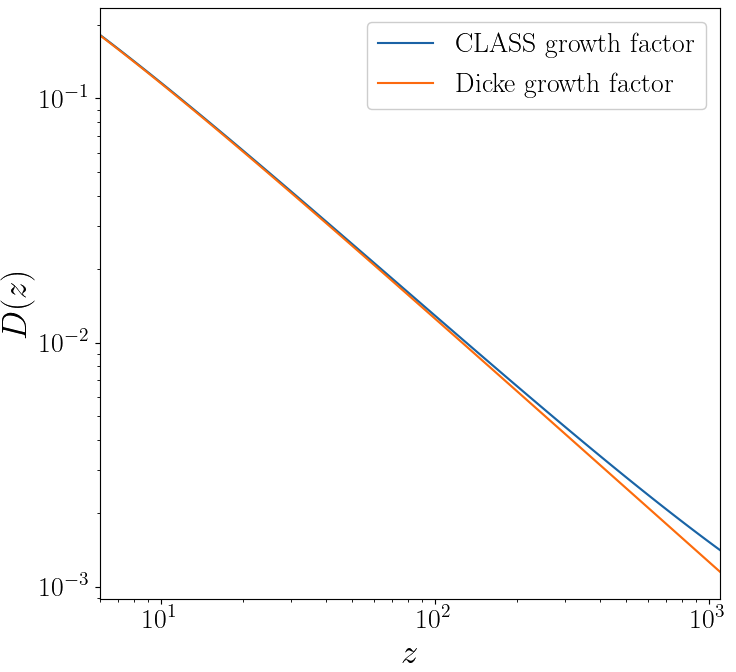}
\caption{Comparison between the {\tt CLASS} growth factor (which solves Eq.~\eqref{eq: A1}) and the Dicke growth factor, as implemented in the standard {\tt 21cmFAST}. At $z=35$ ($z=20$) the relative error is $\sim 1.4\%$ ($\sim 0.98\%$).}
\label{Fig: figure_A1}
\end{figure}

Among its calculations, {\tt CLASS} solves Eq.~\eqref{eq: A1} to find $D\left(z\right)$. In contrast, {\tt 21cmFAST} does not solve Eq.~\eqref{eq: A1}, and instead it adopts the fit of Refs.~\cite{Liddle:1995pd, Carroll:1991mt}, known in the code as the Dicke growth factor. In Fig.~\ref{Fig: figure_A1} we show the two growth factors of the two codes. The agreement between them becomes excellent at low redshifts, although percent-level errors can  still be found for $z\gtrsim20$. At high redshifts, the error of the Dicke fit is no longer negligible and it reaches $\sim20\%$ at $z=1000$. In order to simulate early temperature and ionization fluctuations as precisely as possible (see more on them in Paper II), we therefore had to incorporate the {\tt CLASS} growth factor in {\tt 21cmFirstCLASS}.

\begin{figure}
\includegraphics[width=\columnwidth]{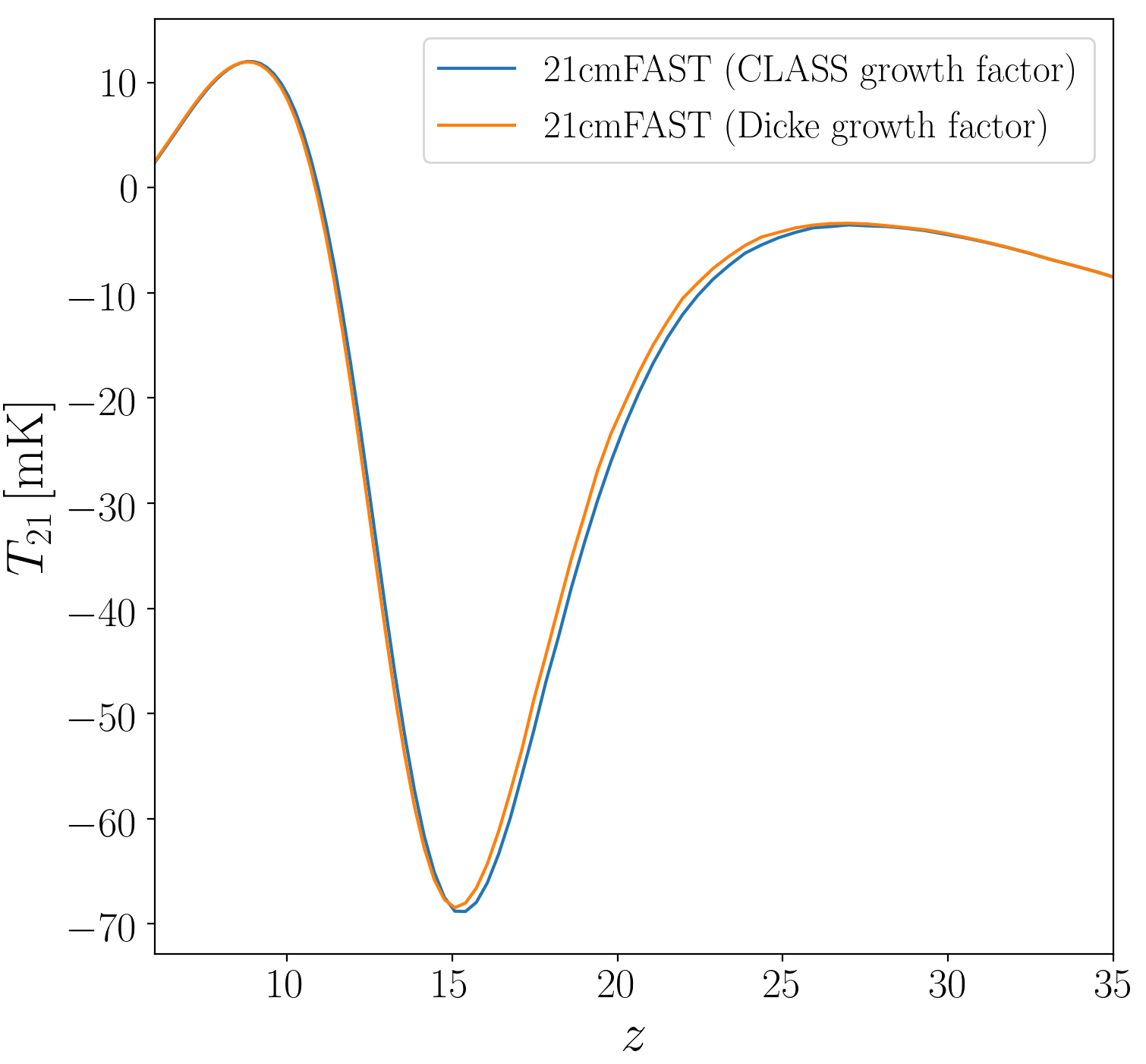}
\caption{Comparison of the 21-cm global signal when different growth factors are considered. In both curves early temperature and ionization fluctuations were discarded by starting the simulation at $z=35$.}
\label{Fig: figure_A2}
\end{figure}

It is interesting however that the small errors of the Dicke fit below $z=35$ can lead to a visible difference in the 21-cm global signal, even if early temperature and ionization fluctuations are discarded, as we show in Fig.~\ref{Fig: figure_A2}. Above $z~\sim27$, the resulting global signal is the same because at this epoch the fluctuations are linear and they cancel each other when the mean of the box is evaluated. Below that redshift, non-linearities become important and the fluctuations (as well as the errors) are no longer canceled. At $z\lesssim15$ the SFRD dominates the fluctuations in the signal, and because the growth factors are nearly the same at that redshift, the Dicke solution to the global brightness temperature coincides with the {\tt CLASS} solution.

\begin{figure}
\includegraphics[width=\columnwidth]{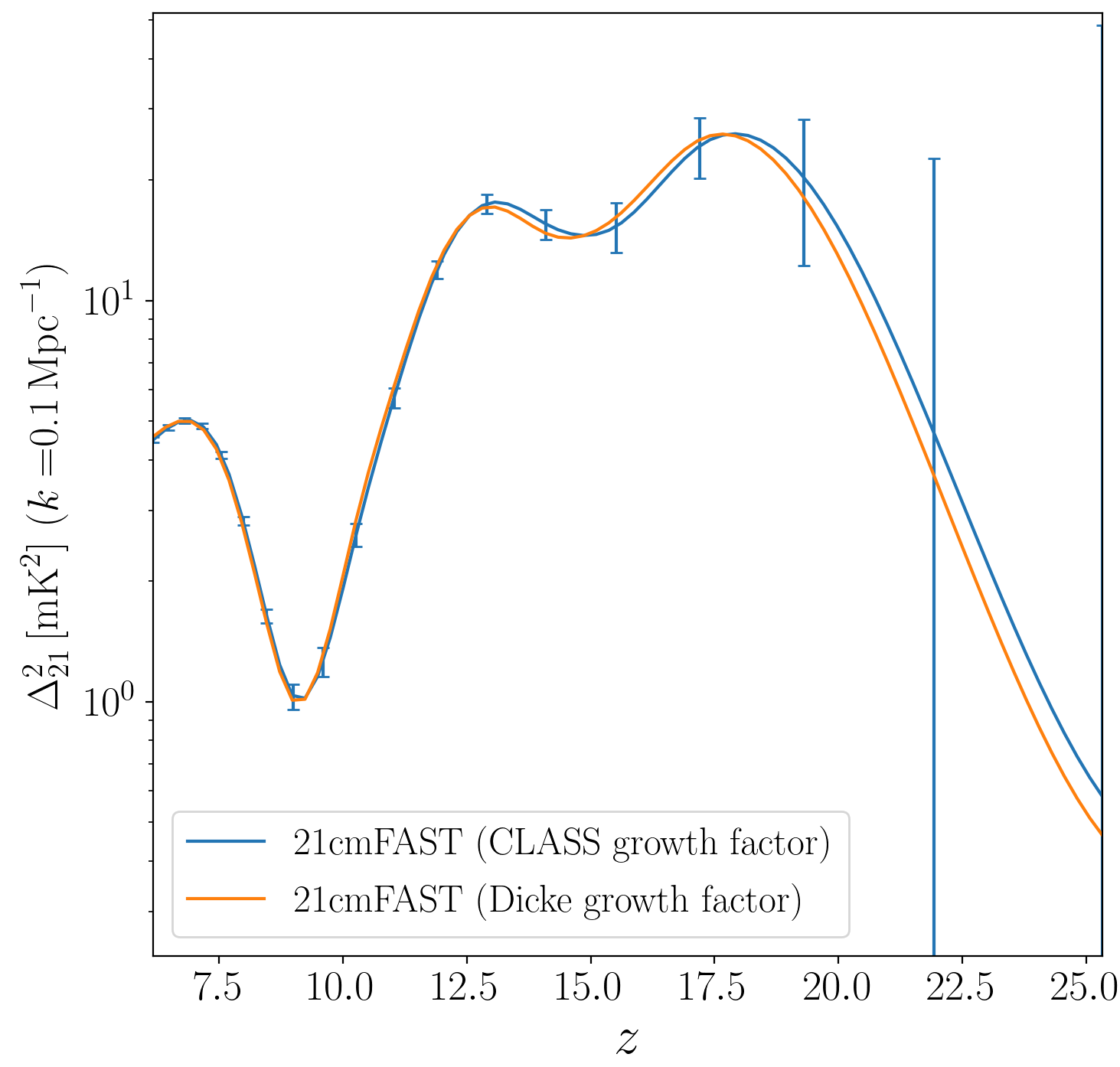}
\caption{Comparison of the 21cm power spectrum when different growth factors are considered. In both curves early temperature and ionization fluctuations were discarded by starting the simulation at $z=35$. Optimistic foreground scenario is assumed for the error bars.}
\label{Fig: figure_A3}
\end{figure}

The errors induced by the Dicke growth factor are enhanced when the 21-cm power spectrum is considered, especially at $z\gtrsim20$, as can be seen in Fig.~\ref{Fig: figure_A3}. Yet, within HERA's range, the errors do not surpass HERA's noise level.

\section{Compton tight coupling approximation}\label{sec: Compton tight coupling approximation}
As was discussed in Sec.~\ref{subsec: 21cmFAST}, above $z=980$ the temperature cannot be evolved precisely if one attempts to solve Eq.~\eqref{eq: 3} numerically via the Euler method but without having a tiny step-size. The reason for this comes from the Compton term in Eq.~\eqref{eq: 3}. At high redshifts this term dominates, leading to $dT_k/dz\propto\left(\Gamma_C/H\right)\left(T_\gamma-T_k\right)$. Since the baryons are tightly coupled to the photons at this epoch, $T_k\to T_\gamma$. However, because $\Gamma_C\gg H$, small initial errors in $T_k$ could cause the solution to overshoot or undershoot $T_\gamma$, depending on the sign of $T_\gamma-T_k$, with oscillations that grow in time. This numerical behavior is well known for interacting fluids in the tight coupling regime. It becomes worse when the temperatures of both fluids have to be simultaneously evolved in time---see Appendix \ref{sec: Dark matter tight coupling approximation}.

To overcome this challenge, many codes use more advanced numerical schemes such as having an adaptive varying step-size or using values from more past samples instead of just the last one. In {\tt 21cmFAST} we cannot use such schemes because the redshift samples (and their corresponding step-sizes) are determined before the evolution of the box begins, and only the last previous box is accessible during the calculation of the current one. Therefore, our numerical scheme is limited to the family of Runge-Kutta solutions. High order Runge-Kutta solutions could increase the required step-size at the price of calculating intermediate redshift samples, but we will see that the simplest lowest-order type of Runge-Kutta solution, namely the Euler method, can be still used without sacrificing valuable computational time.

The trick is to track the difference and the average temperatures of the tightly coupled fluids, instead of tracking the temperatures of the individual fluids. A similar method to the one presented below is already implemented in {\tt CLASS}. To see why such a method is helpful, let us rewrite Eq.~\eqref{eq: 11} (that includes also interaction with SDM) in the following form (note we now denote the kinetic gas temperature with $T_b$ to match our notation in Sec.~\ref{sec: Scattering dark matter})
\begin{equation}\label{eq: B1}
\frac{dT_b}{dz}=\frac{1}{1+z}\left[2T_b-\frac{T_\gamma-T_b}{\epsilon_{\gamma b}}-\frac{2\dot Q_b}{3k_BH}-\frac{1}{H}\left.\frac{dT_b}{dt}\right|_\mathrm{ext}\right],
\end{equation}
where we defined $\epsilon_{\gamma b}\equiv H/\Gamma_C$. Because $T_\gamma\propto\left(1+z\right)$, we also know that
\begin{equation}\label{eq: B2}
\frac{dT_\gamma}{dz}=\frac{T_\gamma}{1+z}.
\end{equation}
When the two fluids are tightly coupled, we approximate $T_b\approx T_\gamma+\mathcal O\left(\epsilon_{\gamma b}\right)$, which is valid as long as $H\ll\Gamma_C$, or $\epsilon_{\gamma b}\ll1$. This is the Compton \emph{tight coupling approximation} or TCA. Within this approximation, we can compare Eqs.~\eqref{eq: B1} and \eqref{eq: B2},
\begin{equation}\label{eq: B3}
T_\gamma=2T_b-\frac{T_\gamma-T_b}{\epsilon_{\gamma b}}-\frac{2\dot Q_b}{3k_BH}-\frac{1}{H}\left.\frac{dT_b}{dt}\right|_\mathrm{ext}+\mathcal O\left(\epsilon_{\gamma b}\right),
\end{equation}
from which we find
\begin{flalign}\label{eq: B4}
\nonumber&\Delta T_{\gamma b}\equiv T_\gamma-T_b&
\\&\hspace{7mm}=\epsilon_{\gamma b}\left[2T_b-T_\gamma-\frac{2\dot Q_b}{3k_BH}-\frac{1}{H}\left.\frac{dT_b}{dt}\right|_\mathrm{ext}\right]+\mathcal O\left(\epsilon_{\gamma b}^2\right).&
\end{flalign}
Furthermore, by adding Eqs.~\eqref{eq: B1} and \eqref{eq: B2} we can find a differential equation for $\bar T_{\gamma b}\equiv\left(T_\gamma+T_b\right)/2$,
\begin{flalign}\label{eq: B5}
\nonumber&\frac{d\bar T_{\gamma b}}{dz}=\frac{1}{2\left(1+z\right)}\Bigg[T_\gamma+2T_b-\frac{T_\gamma-T_b}{\epsilon_{\gamma b}}&
\\&\hspace{10mm}-\frac{2\dot Q_b}{3k_BH}-\frac{1}{H}\left.\frac{dT_b}{dt}\right|_\mathrm{ext}\Bigg]=\frac{T_\gamma}{1+z}+\mathcal O\left(\epsilon_{\gamma b}\right),&
\end{flalign}
where the second line follows Eq.~\eqref{eq: B3}. Not surprisingly, we see that the average temperature of the tightly coupled baryon-photon fluid follows the CMB temperature. We would need a second differential equation, for the temperature difference $\Delta T_{\gamma b}$. According to Eq.~\eqref{eq: B4}, this is equivalent to finding a differential equation for $\epsilon_{\gamma b}$. From Eqs.~\eqref{eq: 4} and \eqref{eq: B2}, a simple calculation yields
\begin{equation}\label{eq: B6}
\frac{d\epsilon_{\gamma b}}{dz}=\epsilon_{\gamma b}\left(\frac{1}{H}\frac{dH}{dz}-\frac{1}{x_e\left(1+x_e\right)}\frac{dx_e}{dz}-\frac{4}{1+z}\right).
\end{equation}

Eqs.~\eqref{eq: B4}-\eqref{eq: B6} are the Compton-TCA equations. Unlike Eq.~\eqref{eq: 3} or Eq.~\eqref{eq: 11}, they do not contain terms that approach zero or infinity in the strong coupling limit and they are thus numerically more stable. The strategy in our code for solving for $T_b$ is as follows:
\begin{enumerate}
\item At each step, we calculate $\epsilon_{\gamma b}\equiv H/\Gamma_C$. If $\epsilon_{\gamma b}>\epsilon_{\gamma b}^{\mathrm{th}}$, where $\epsilon_{\gamma b}^{\mathrm{th}}$ is some threshold value, the TCA does not have to be applied, and we solve Eq.~\eqref{eq: 3}.
\item Otherwise, we compute $\bar T_{\gamma b}$, and evolve $\bar T_{\gamma b}$ and $\epsilon_{\gamma b}$ via Eqs.~\eqref{eq: B5} and \eqref{eq: B6}, respectively.
\item We then compute the current $\Delta T_{\gamma b}$ via Eq.~\eqref{eq: B4}.
\item Finally, we find the current gas temperature with $T_b=\bar T_{\gamma b}-\Delta T_{\gamma b}/2$.
\end{enumerate}
With this prescription, we can run {\tt 21cmFirstCLASS} with a constant $\Delta z_n=0.1$ from recombination to $z=35$ and thus reduce the total amount of redshift samples by $\sim8000$. In our code we have set $\epsilon_{\gamma b}^{\mathrm{th}}=5\times10^{-5}$ since this choice corresponds to $\epsilon_{\gamma b}\left(z=980\right)\approx\epsilon_{\gamma b}^{\mathrm{th}}$, though we comment that $\epsilon_{\gamma b}^{\mathrm{th}}$ can be even three orders of magnitude greater and the desired evolution would be still obtained.

\section{Dark matter tight coupling approximation}\label{sec: Dark matter tight coupling approximation}
A similar problem to the one discussed in Appendix \ref{sec: Compton tight coupling approximation} happens when baryons interact with SDM. According to Eq.~\eqref{eq: 11}-\eqref{eq: 12}, the changes in $T_b$ and $T_\chi$ depend on $\dot Q_b$ and $\dot Q_\chi$, but according to Eq.~\eqref{eq: 18}-\eqref{eq: 19}, these quantities depend on the difference between $T_b$ and $T_\chi$. If $\Gamma_{\chi b}\gg H$ (or $\left(n_b/n_\chi\right)\Gamma_{\chi b}\gg H$), then $T_b-T_\chi\to0$ and small numerical deviations from the true solution will cause the error to diverge, in both fluids. Moreover, it becomes unclear what happens in a scenario where the fluids are tightly coupled, but only one of them is strongly affected by an external source, e.g. X-rays that heat-up only the baryons fluid. This is why the DM-TCA algorithm that we derive below does not serve only as a means to reduce runtime, but in fact it is \emph{indispensable} to get the right evolution at low redshifts.

We begin by rewriting Eq.~\eqref{eq: 11}-\eqref{eq: 12} in the following form,
\begin{eqnarray}\label{eq: C1}
\nonumber\frac{1}{H}\frac{dT_b}{dt}&=&-2T_b+\frac{T_\gamma-T_b}{\epsilon_{\gamma b}}+\frac{T_\chi-T_b}{\epsilon_b}+\frac{1}{H}\left.\frac{dT_b}{dt}\right|_\mathrm{ext}
\\&&+\frac{2}{3k_B}\frac{\rho_\chi}{\rho_\mathrm{tot}}\frac{V_{\chi b}}{H}\sum_t\frac{m_\chi m_b}{m_\chi+m_t}D_t\left(V_{\chi b}\right),
\end{eqnarray}
\begin{eqnarray}\label{eq: C2}
\nonumber\frac{1}{H}\frac{dT_\chi}{dt}&=&-2T_\chi+\frac{T_b-T_\chi}{\epsilon_\chi}+\frac{1}{H}\left.\frac{dT_\chi}{dt}\right|_\mathrm{ext}
\\&&+\frac{2}{3k_B}\frac{\rho_b}{\rho_\mathrm{tot}}\frac{V_{\chi b}}{H}\sum_t\frac{m_\chi m_t}{m_\chi+m_t}D_t\left(V_{\chi b}\right),
\end{eqnarray}
where we have defined the DM-TCA small parameters,
\begin{equation}\label{eq: C3}
\epsilon_b\equiv\frac{H}{\Gamma_{\chi b}},\qquad\epsilon_\chi\equiv\frac{n_\chi}{n_b}\epsilon_b.
\end{equation}
It will be convenient to define a symmetrized small parameter,
\begin{flalign}\label{eq: C4}
\nonumber&\epsilon_{\chi b}\equiv\frac{n_\chi}{n_\chi+n_b}\epsilon_b=\frac{n_b}{n_\chi+n_b}\epsilon_\chi&
\\&=\sqrt{\frac{\pi}{8}}\frac{3H}{\sigma_{-4}c^4\left(n_\chi+n_b\right)m_\chi}\left[\sum_t\frac{n_t}{n_b}\frac{m_t\mathrm{e}^{-r_t^2/2}}{\left(m_t+m_\chi\right)^2u_{\chi t}^3}\right]^{-1}.&
\end{flalign}

In the DM-TCA we have $\epsilon_{\chi b}\ll1$ and $T_b=T_\chi+\mathcal O\left(\epsilon_{\chi b}\right)$. This allows us to compare Eqs.~\eqref{eq: C1} and \eqref{eq: C2}, and find that in the strong coupling limit the temperature difference is
\begin{flalign}\label{eq: C5}
\nonumber&\Delta T_{b\chi}\equiv T_b-T_\chi=\epsilon_{\chi b}\Bigg[\frac{T_\gamma-T_b}{\epsilon_{\gamma b}}+\frac{1}{H}\left.\frac{dT_b}{dt}\right|_\mathrm{ext}-\frac{1}{H}\left.\frac{dT_\chi}{dt}\right|_\mathrm{ext}&
\\&\hspace{10mm}-\frac{2}{3k_B}\frac{V_{\chi b}m_\chi}{H\rho_\mathrm{tot}}\sum_t\frac{\rho_bm_t-\rho_\chi m_b}{m_\chi+m_t}D_t\left(V_{\chi b}\right)\Bigg]+\mathcal O\left(\epsilon_{\chi b}^2\right).&
\end{flalign}
By adding together Eqs.~\eqref{eq: C1} and \eqref{eq: C2}, and using Eq.~\eqref{eq: C5}, we can also find a differential equation for $\bar T_{\chi b}\equiv\left(T_b+T_\chi\right)/2$,
\begin{flalign}\label{eq: C6}
\nonumber&\frac{d\bar T_{\chi b}}{dz}=\frac{1}{1+z}\Bigg[2\bar T_{\chi b}-\frac{n_b}{n_b+n_\chi}\left(\frac{T_\gamma-T_b}{\epsilon_{\gamma b}}+\frac{1}{H}\left.\frac{dT_b}{dt}\right|_\mathrm{ext}\right)&
\\\nonumber&\hspace{7mm}-\frac{n_\chi}{n_b+n_\chi}\frac{1}{H}\left.\frac{dT_\chi}{dt}\right|_\mathrm{ext}-\frac{1}{n_b+n_\chi}\frac{2}{3k_B}\frac{\rho_b\rho_\chi}{H\rho_\mathrm{tot}}V_{\chi b}D\left(V_{\chi b}\right)\Bigg]&
\\&\hspace{7mm}+\mathcal O\left(\epsilon_{\chi b}\right).&
\end{flalign}
To solve for $T_b$ and $T_\chi$ we require another equation for $\Delta T_{b\chi}$. According to Eq.~\eqref{eq: C5}, this is equivalent to having an equation for $\epsilon_{\chi b}$. Since $n_b\propto n_\chi\propto\left(1+z\right)^3$, then from Eqs.~\eqref{eq: C3} and \eqref{eq: C4} we have
\begin{equation}\label{eq: C7}
\frac{d\epsilon_{\chi b}}{dz}=\frac{\epsilon_{\chi b}}{\epsilon_b}\frac{d\epsilon_b}{dz}=\epsilon_{\chi b}\left(\frac{1}{H}\frac{dH}{dz}-\frac{1}{\Gamma_{\chi b}}\frac{d\Gamma_{\chi b}}{dz}\right),
\end{equation}
where the derivative of the energy transfer rate $\Gamma_{\chi b}$ can be evaluated from its definition, Eq.~\eqref{eq: 20}, up to $\mathcal O\left(\epsilon_{\chi b}\right)$ corrections,
\begin{flalign}\label{eq: C8}
&\nonumber\frac{d\Gamma_{\chi b}}{dz}=\sqrt{\frac{2}{\pi}}\frac{2\sigma_{-4}c^4\rho_\chi}{3n_b}\sum_t\Bigg\{\frac{\rho_t\mathrm{e}^{-r_t^2/2}}{\left(m_t+m_\chi\right)^2u_{\chi t}^3}\times&
\\&\hspace{8mm}\left[\frac{3}{1+z}-\frac{r_t}{u_{\chi t}}\frac{dV_{\chi b}}{dz}-\left(3-r_t^2\right)\frac{m_t+m_\chi}{m_tm_\chi}\frac{k_B}{u_{\chi t}^2}\frac{d\bar T_{\chi b}}{dz}\right]\Bigg\}.&
\end{flalign}
Note that $d\Gamma_{\chi b}/dz\propto\Gamma_{\chi b}$ in the special case in which the SDM interacts with a single type of particles.

Eqs.~\eqref{eq: C5}-\eqref{eq: C8} are the DM-TCA equations. It is crucial to understand that if $\epsilon_{\chi b}\ll 1$, that does not guarantee that both $\epsilon_b$ and $\epsilon_\chi$ are much smaller than unity. This is because $\epsilon_b\propto\rho_\chi^{-1}$ and $\epsilon_\chi\propto\rho_t^{-1}$. So for example, if $f_\chi \ll 1$ such that $\epsilon_b\gg1$, then the baryons are not tightly coupled to the SDM. However, if $\sigma_{-4}$ is large enough, then even if $f_\chi\ll1$, it might be that $\epsilon_\chi\ll1$ and the SDM is coupled to the baryons. This is similar to the early coupling between baryons and CMB photons; the latter outnumber the former, and thus the baryons are tightly coupled to the CMB, while the CMB photons are insensitive to the baryons. Since $\epsilon_{\chi b}\leq\epsilon_\chi,\epsilon_b$, if either $\epsilon_b\ll1$ or $\epsilon_\chi\ll1$, that implies that $\epsilon_{\chi b}\ll1$, and we can evolve $\bar T_{\chi b}$ and $\Delta T_{b\chi}$ with the DM-TCA equations we formulated above.

All of these considerations have been implemented in {\tt 21cmFirstCLASS}. Below we present the algorithm we use in our code to solve for $T_b$ and $T_\chi$.
\begin{enumerate}
\item We begin by calculating $\epsilon_b$ and $\epsilon_\chi$ via Eq.~\eqref{eq: C3}. If at least one of them is smaller than the threshold $\epsilon_{\chi b}^\mathrm{th}$, we use the DM-TCA equations, Eqs.~\eqref{eq: C5}-\eqref{eq: C8}, to find the updated values of $\bar T_{\chi b}$ and $\Delta T_{b\chi}$.
\begin{enumerate}
\item When we use the DM-TCA equations, we check if $\epsilon_{\gamma b}<\epsilon_{\gamma b}^\mathrm{th}$. If this condition is satisfied, we are in the special scenario where the three fluids (baryons, SDM and CMB photons) are strongly coupled. In this case we evaluate $\left(T_\gamma-T_b\right)/\epsilon_{\gamma b}$ that appears in Eqs.~\eqref{eq: C5}-\eqref{eq: C6} with the Compton-TCA Eq.~\eqref{eq: B4}.
\end{enumerate}
\item Next, we solve for the baryons temperature $T_b$.
\begin{enumerate}
\item We check if $\epsilon_{\gamma b}>\epsilon_{\gamma b}^\mathrm{th}$ and $\epsilon_{b}>\epsilon_{\chi b}^\mathrm{th}$. If these two conditions are satisfied, or alternatively $dT_b/dt|_\mathrm{ext}>2\dot Q_b/\left(3k_B\right)$, we solve the usual differential equation for $T_b$, Eq.~\eqref{eq: 11}. The latter condition reflects the understanding that the baryons cannot be tightly coupled to the SDM if an external heating source, such as X-rays, is more dominant.
\item Otherwise, if $\epsilon_{\gamma b}\leq\epsilon_{\gamma b}^\mathrm{th}$, we use the Compton-TCA equations, Eqs.~\eqref{eq: B4}-\eqref{eq: B6}, to solve for $T_b$. This reflects our assumption that the coupling of the baryons with the SDM cannot be stronger than the coupling of the baryons with the CMB. Cross-sections that break this assumption imply that $T_b\neq T_\gamma$ at recombination and have been ruled out by CMB observations.
\item Otherwise, the baryons are not tightly coupled to the CMB, but they are tightly coupled to the SDM, and we can use $\bar T_{\chi b}$ and $\Delta T_{b\chi}$ that we obtained in item 1 to find $T_b$ via $T_b=\bar T_{\chi b}+\Delta T_{b\chi}/2$.
\end{enumerate}
\item Finally, we solve for the SDM temperature $T_\chi$.
\begin{enumerate}
\item We check if $\epsilon_{\chi}>\epsilon_{\chi b}^\mathrm{th}$ or if $dT_\chi/dt|_\mathrm{ext}>2\dot Q_\chi/\left(3k_B\right)$ (the latter condition can be satisfied at low redshifts, when the clustering of SDM becomes important). If one of these conditions is satisfied, we solve the usual differential equation for $T_\chi$, Eq.~\eqref{eq: 12}.
\item Otherwise, the SDM is tightly coupled to the baryons, and we can use $\bar T_{\chi b}$ and $\Delta T_{b\chi}$ that we obtained in item 1 to find $T_\chi$ via $T_\chi=\bar T_{\chi b}-\Delta T_{b\chi}/2$.
\end{enumerate}
\end{enumerate}

For the threshold of the DM-TCA small parameter we use $\epsilon_{\chi b}^\mathrm{th}=\epsilon_{\gamma b}^\mathrm{th}=5\times10^{-5}$ at high redshifts ($z>100$) and $\epsilon_{\chi b}^\mathrm{th}=10^{-2}$ at low redshifts. We have confirmed that the results of our code are insensitive to these particular values. Moreover, we have confirmed the correctness of our solutions to $T_b$ and $T_\chi$ by comparing them to the solutions that can be obtained by solving Eqs.~\eqref{eq: 11}-\eqref{eq: 14} with {\tt Mathematica}~\cite{Mathematica} (when all the fluctuations in the box are turned off and we set $L_X=35$, namely no X-ray heating). Unlike our code, {\tt Mathematica} solves differential equations by adjusting the step-size so that the estimated error in the solution is just within the specified absolute and relative tolerances. In fact, our DM-TCA algorithm presented above allows solving correctly the differential equations even for cross-sections that are large enough such that the normal settings in {\tt Mathematica} fail to solve the equations.

\section{Small temperature correction for $S_\alpha$}\label{sec: Small temperature correction for Salpha}

The Ly$\alpha$ coupling in Eq.~\eqref{eq: 2} is given by~\cite{Mittal:2020kjs}
\begin{equation}\label{eq: D1}
\tilde x_\alpha=\frac{J_\alpha}{J_0}\tilde S_\alpha,
\end{equation}
where $J_\alpha$ is the Ly$\alpha$ flux, and $J_0$ is
\begin{eqnarray}\label{eq: D2}
\nonumber J_0&=&\frac{9A_{10}T_\gamma}{8\pi\lambda^2_\mathrm{Ly\alpha}\gamma_\alpha T_\star}
\\&=&5.54\times10^{-12}\left(1+z\right)\mathrm{cm^{-2}\,sec^{-1}\,Hz^{-1}\,sr^{-1}},
\end{eqnarray}
where $\lambda^2_\mathrm{Ly\alpha}=121.567\,\mathrm{nm}$ is the Ly$\alpha$ frequency, $A_{10}=2.85\times10^{-15}\,\mathrm{sec}^{-1}$ is the spontaneous emission coefficient from the excited hyperfine level to the ground state, and $\gamma_\alpha\approx50\,\mathrm{MHz}$ is the half width at half maximum of the Ly$\alpha$ resonance line.

The quantity $\tilde S_\alpha$ that appears in Eq.~\eqref{eq: D1} is a correction to $\tilde x_\alpha$ due to spectral distortions. In order to find it, one must solve the steady-state Fokker-Planck equation. This was first numerically solved by Chen \& Miralde-Escud\'{e}~\cite{Chen:2003gc} and later was refined by Hirata~\cite{Hirata:2005mz}, who found a complicated fit for $\tilde S_\alpha$ that depends on $T_k$, $T_s$ and the Gunn-Peterson optical depth $\tau_\mathrm{GP}$. In addition, Hirata found a fit for the color temperature, $T_\alpha^{-1}=T_k^{-1}+T_\mathrm{se}T_k^{-1}\left(T_s^{-1}-T_k^{-1}\right)$, where $T_\mathrm{se}$ accounts for the correction in the color temperature due spin exchange and is given by
\begin{equation}\label{eq: D3}
T_\mathrm{se}=\left(\frac{\lambda_\mathrm{Ly\alpha}}{\lambda_{21}}\right)^2\frac{m_\mathrm{H}c^2}{9k_B}\approx0.4\,\mathrm{K},
\end{equation}
where $\lambda_{21}\approx21\,\mathrm{cm}$ is the wavelength of a 21-cm photon. The fits discovered by Hirata are implemented in {\tt 21cmFAST}.

\begin{figure}
\includegraphics[width=\columnwidth]{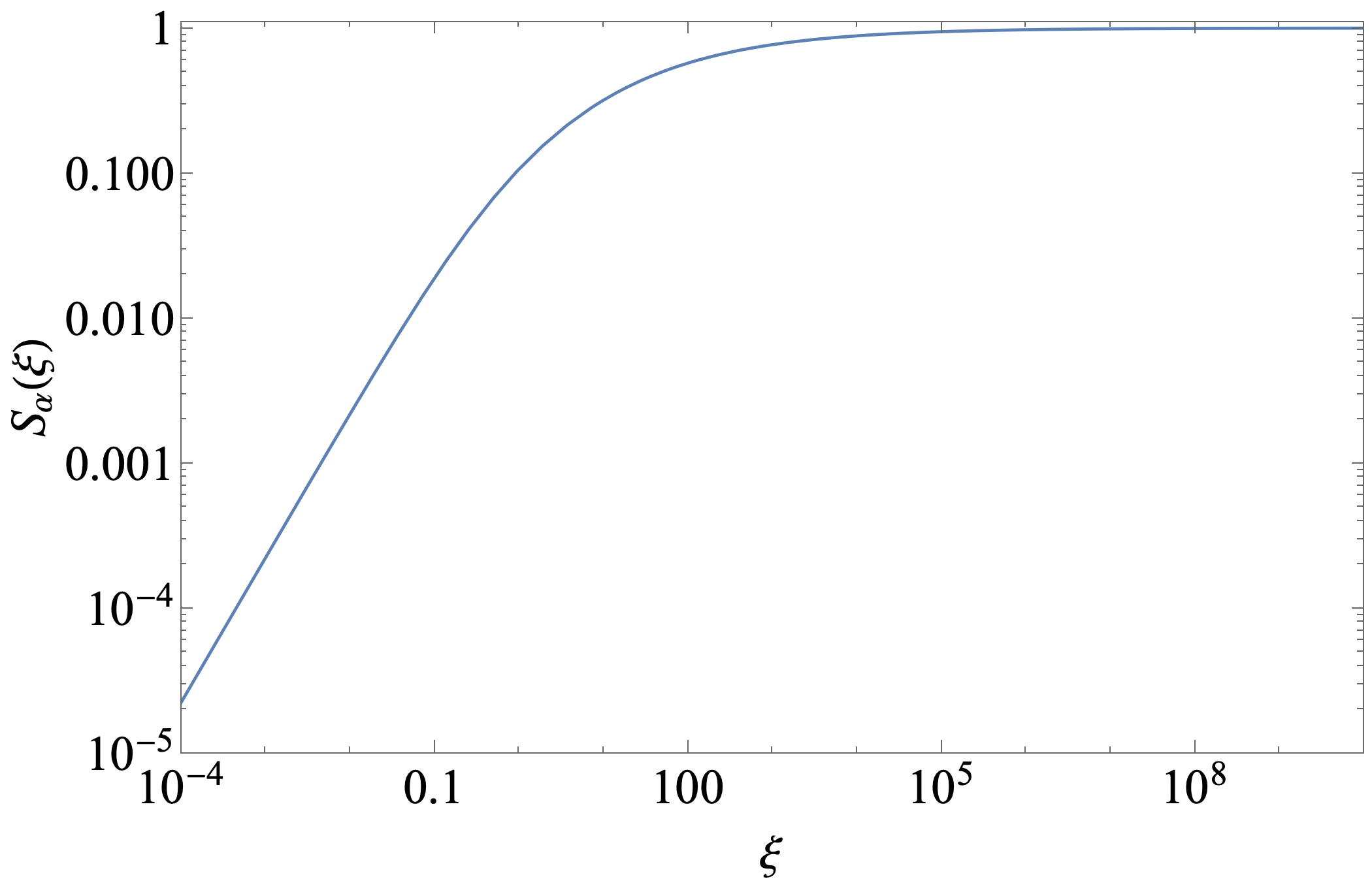}
\caption{The function $\tilde S_\alpha\left(\xi\right)$ as given by Eq.~\eqref{eq: D5}.}
\label{Fig: figure_D1}
\end{figure}

\begin{figure}
\includegraphics[width=\columnwidth]{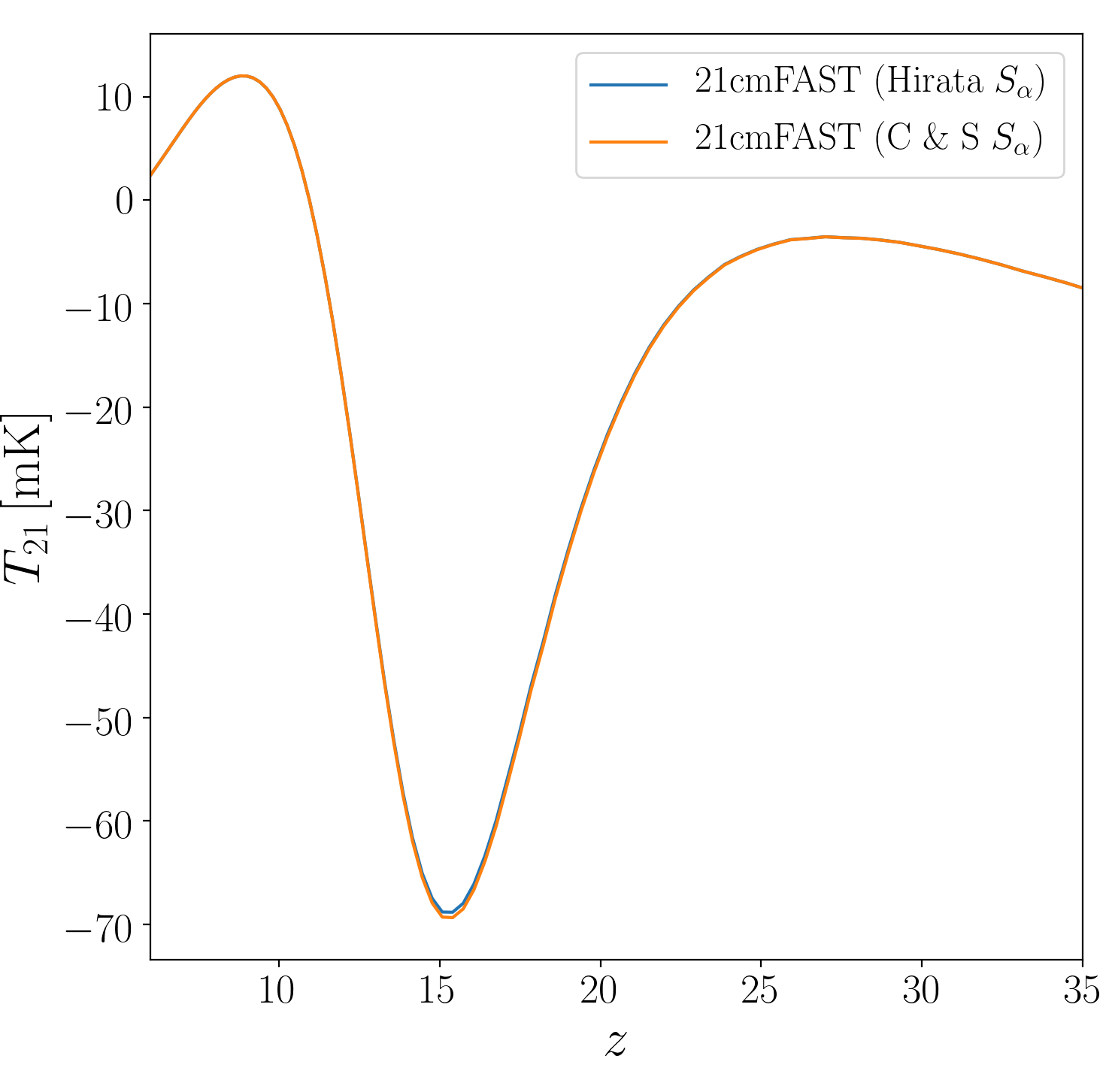}
\caption{Comparison of the 21cm global signal when $\tilde S_\alpha$ is calculated from Ref.~\cite{Hirata:2005mz} (Hirata) or Ref.~\cite{Chuzhoy:2005wv} (Chuzhoy \& Shapiro). In both curves early temperature and ionization fluctuations were discarded by starting the simulation at $z=35$.}\label{Fig: figure_D2}
\end{figure}

\begin{figure}
\includegraphics[width=\columnwidth]{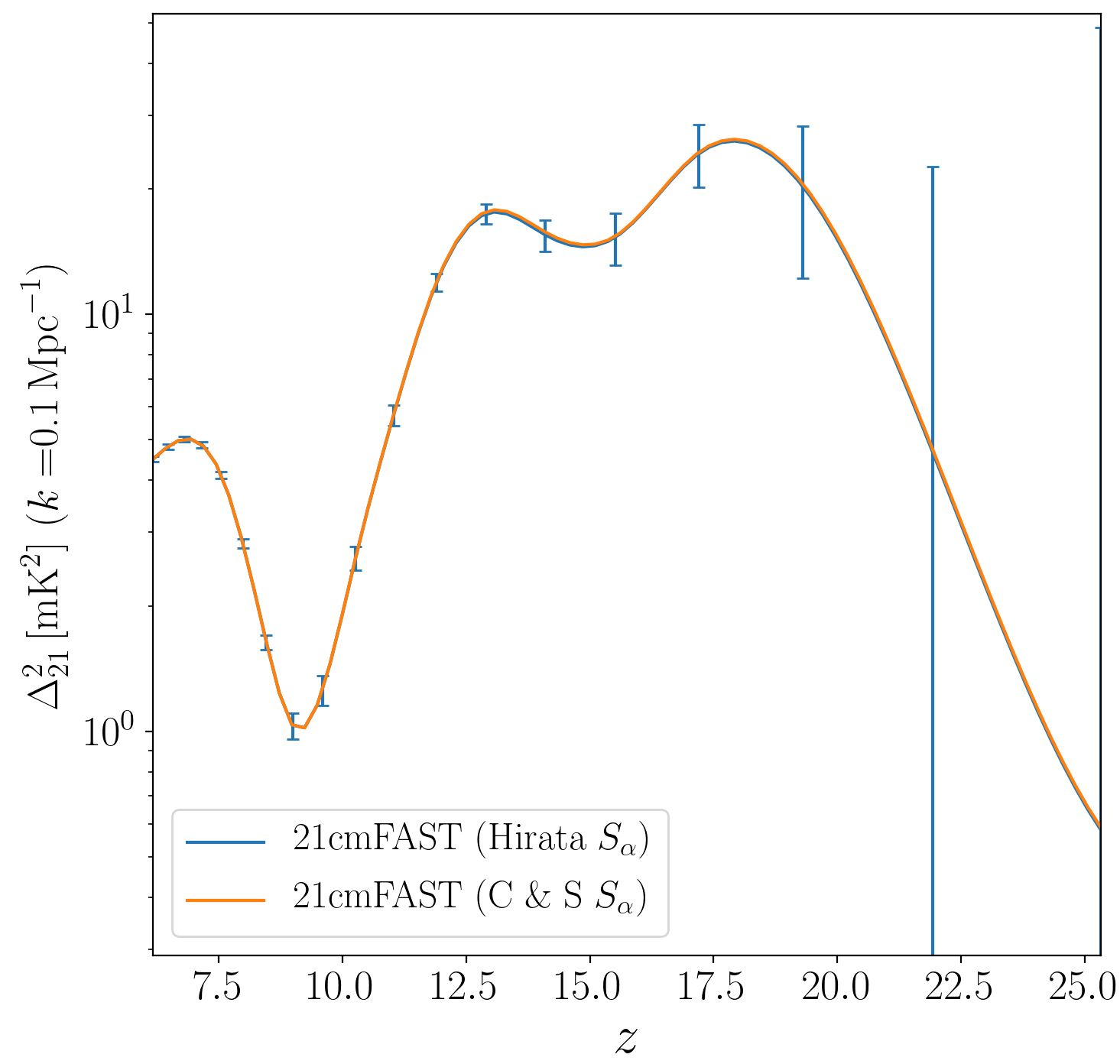}
\caption{Comparison of the 21cm power spectrum when $\tilde S_\alpha$ is calculated from Ref.~\cite{Hirata:2005mz} (Hirata) or Ref.~\cite{Chuzhoy:2005wv} (Chuzhoy \& Shapiro). In both curves early temperature and ionization fluctuations were discarded by starting the simulation at $z=35$. Optimistic foreground scenario is assumed for the error bars.}
\label{Fig: figure_D3}
\end{figure}

Shortly after Hirata's work, Chuzhoy \& Shapiro~\cite{Chuzhoy:2005wv} found an analytical solution to the steady-state Fokker-Planck equation by approximating the spectrum with the absorption profile appropriate to Lorentzian wings (this was first done by Grachev~\cite{1989Afz....30..347G}). This is known as the wing approximation. Based on their work, Furlanetto \& Pritchard~\cite{Furlanetto:2006fs} gave analytical estimates, including for the color temperature, 
\begin{equation}\label{eq: D4}
T_\alpha=T_s\,\frac{T_k+T_\mathrm{se}}{T_s+T_\mathrm{se}}.
\end{equation}
Note that in the limit $T_\mathrm{se}\ll T_k, T_s$ Eq.~\eqref{eq: D4} converges to Hirata's fit. Recently, Ref.~\cite{Mittal:2020kjs} used the results of Furlanetto \& Pritchard to write the analytical estimate for $\tilde S_\alpha$ in the following way,
\begin{equation}\label{eq: D5}
\tilde S_\alpha\left(\xi\right) = 1-\int_0^\infty\mathrm{e}^{-\xi\left(u/3\right)^3}\mathrm{e}^{-u}\,du=\begin{cases}
1 & \xi\to\infty \\
\frac{2}{9}\xi & \xi\ll 1
\end{cases},
\end{equation}
where
\begin{eqnarray}\label{eq: D6}
\nonumber\xi&\equiv&\frac{3\nu_\mathrm{Ly\alpha}m_\mathrm{H}H\left(k_BT_k\right)^2}{\pi A_\alpha\gamma_\alpha c \hbar^3n_\mathrm{H}\left(1-x_e\right)}
\\\nonumber&\approx&760\left(\frac{\Omega_{m}h^{2}}{0.143}\right)^{1/2}\left(\frac{\Omega_{b}h^{2}}{0.0223}\right)^{-1}\left(\frac{1-\mathrm{Y_{He}}}{0.755}\right)^{-1}
\\&&\times\left(\frac{T_{k}}{10\,\mathrm{K}}\right)^{2}\left(\frac{1+z}{15}\right)^{-3/2}\frac{1}{\left(1+\delta_{b}\right)\left(1-x_{e}\right)},
\end{eqnarray}
where $\nu_\mathrm{Ly\alpha}=2.47\times10^{15}\,\mathrm{Hz}$ the Ly$\alpha$ frequency and $A_\alpha=6.25\times10^8\,\mathrm{Hz}$ the spontaneous Ly$\alpha$ emission coefficient. The fiducial values in Eq.~\eqref{eq: D6} correspond to typical values of $T_k$ at $z=15$ in $\Lambda$CDM (c.f. Fig.~\ref{Fig: figure_1}). According to Fig.~\ref{Fig: figure_D1}, $\xi\sim10^3$ corresponds to $\tilde S_\alpha\sim0.5$.

In SDM however, the gas kinetic temperature may reach very low temperatures (c.f. Fig.~\ref{Fig: figure_9}) where Hirata's fit to $\tilde S_\alpha$ no longer works. Therefore, for our SDM calculations in this paper we follow\footnote{It is worth mentioning that another fit for $\tilde S_\alpha$ at low temperature exists in the literature~\cite{Barkana:2022hko, Barkana:2016nyr}. However, implementing this fit in {\tt 21cmFirstCLASS} did not yield results that agree with either \cite{Hirata:2005mz} or \cite{Chuzhoy:2005wv}.} ~\cite{Driskell:2022pax} and work with Eqs.~\eqref{eq: D4}-\eqref{eq: D6}. In Figs.~\ref{Fig: figure_D2} and \ref{Fig: figure_D3}  we verify that in $\Lambda$CDM the output of our code is not sensitive to the chosen method. Indeed, there is an excellent agreement between the two methods.

Because $\xi\propto T_k^2$ and $T_k$ can be smaller by a factor of $\sim10^2$ compared to $\Lambda$CDM, the value of $\xi$ may drop below $0.1$ and from Fig.~\ref{Fig: figure_D1} it is implied that $\tilde S_\alpha<0.01$. This explains why in Fig.~\ref{Fig: figure_9}, $T_s$ cannot reach very low temperatures even though $T_k$ does.

\newpage
\bibliography{21cmFirstCLASS_I_cosmological_tool_and_SDM.bib}

\end{document}